\documentclass[12pt]{iopart}
\usepackage{iopams}  
\usepackage{graphicx} 
\DeclareUnicodeCharacter{2212}{-}
\begin{document}

\title{Route to chaos in two-dimensional discrete parametric maps with bistable potentials}

\author{Alain M. Dikand\'e}

\address{Laboratory of Research on Advanced Materials and Nonlinear Science (LaRAMaNS), Department of Physics, Faculty of Science, University of Buea P.O. Box 63 Buea, Cameroon.}
\ead{dikande.alain@ubuea.cm}
\vspace{10pt}
\begin{indented}
\item[]March 2021
\end{indented}

\begin{abstract}
The texture of phase space and bifurcation diagrams of two-dimensional discrete maps describing a lattice of interacting oscillators, confined in on-site potentials with deformable double-well shapes, are examined. The two double-well potentials considered belong to a family proposed by Dikand\'e and Kofan\'e (A. M. Dikand\'e and T. C. Kofan\'e, Solid State Commun. vol. 89, p. 559, 1994), whose shapes can be tuned distinctively: one has a variable barrier height and the other has variable minima positions. However the two parametrized double-well potentials reduce to the $\phi^4$ substrate, familiar in the studies of structural phase transitions in centro-symmetric crystals or bistable processes in biophysics. It is shown that although the parametric maps are area preserving their routes to chaos display different characteristic features: the first map exhibits a cascade of period-doubling bifurcations with respect to the potential amplitude, but period-halving bifurcations with respect to the shape deformability parameter. On the other hand the first bifurcation of the second map always coincides with the first pitchfork bifurcation of the $\phi^4$ map. However, an increase of the deformability parameter shrinks the region between successive period-doubling bifurcations. The two opposite bifurcation cascades characterizing the first map, and the shrinkage of regions between successive bifurcation cascades which is characteristic of the second map, suggest a non-universal character of the Feigenbaum-number sequences associate with the two discrete parametric double-well maps.
\end{abstract}

%
%
%
%
%

\section{\label{sec:level1}Introduction}
\begin{figure}\centering 
 \includegraphics[width=3.in,height=1.8in]{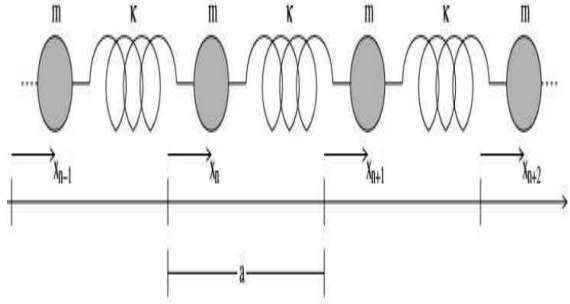}
 \caption{\label{figa}(Color online). Sketch of a $1D$ chain of oscillators interacting via identical springs. $X_n$ denotes the displacement of the $n^{th}$ oscillator with respect to the equilibrium.} 
 \end{figure}

Real physical systems exhibit extremely complex structures, and consequently a broad range of unexpected behaviors due to the collective motion of their constituants mediated by competing effects of many-body and one-body interactions. Fig. \ref{figa} represents a  discrete chain of identical atoms of mass $m$, interacting via springs of a spring constant $K$.
The figure is representative of a large variety of physical systems observed in nature, ranging from solid-state physics to chemistry, biophysics and chemistry \cite{toda,remoi,chua,pana}. In these physical systems fig. \ref{figa} provides simple representation of the lattice structures of atomic or molecular crystals, the structures of organized cellular populations in tissues and embryonic cells, the organization of glycolitic oscillations in suspensions of yeast cells in unison, the structure of an assembly of pacemaker cells in the sino-atrial node, to name these few physical contexts~\cite{chua}. Although they share in common the mutual interactions of their elementary constituants (e.g. atoms, molecules, cells in tissues, chemical species, etc.), every physical system possesses a unique feature that stems from feedback functions characterizing the system \cite{chua}. These feedback functions are governed by one-body forces which are typically nonlinear, in the specific context of Hamiltonian systems they are conservative forces that derive from one-body potentials. \par 
Several potentials are found in the literature, each of them resulting from efforts to provide a realistic description of a specific physical process. But the two most common types of potential used in nonlinear Hamiltonian dynamics are the periodic potentials with infinitely degenerate extrema, and piecewise potentials with finite numbers of extrema. Of the first type the sine-Gordon potential is best known and probably the most studied \cite{sin2,sin3}, it was proposed to model systems in which ordered states are governed by periodic forces. This is for instance the case of the emergence of incommensurate orders due to dislocations in Frenkel-Kontorova crystals \cite{fren,dikan}, commensurate and incommensurate orders in chiral liquid crystals \cite{lcrys}, the incommensurate charge-density-wave order mediated by phasons in low-dimensional conductors \cite{com}, the formation and propagation of fluxons in Josephson-Junction superconductors \cite{kiv}, the equilibrium configuration of finite masses suspended on periodically aligned coupled pendula, etc. As for potentials with finite extrema, polynomial-type potentials are most popular among which the canonical $\phi^4$ potential \cite{krum,bak2,bak1}. It was introduced initially in the study of ferroelectric transitions in one-dimensional ($1D$) centro-symmetric crystals such as perovskites, and later on generalized to second-order phase transitions \cite{krum,fer1,fer2}. In this last context the $\phi^4$ potential stands for the Ginzburg-Landau energy functional, whose two degenerate minima separated by a potential barrier is assumed to  mimic orderings of atoms/molecules in crystals with an inversion (or point) symmetry \cite{fer1,fer2}. But the applicability of the $\phi^4$ potential goes beyond second-order phase transitions in $1D$ crystals, in fact the $\phi^4$ model is a paradigm for bistable systems which can be found in biophysics \cite{bio1,bio2}, chemistry \cite{bio2}, economy \cite{bio2,econ1}, population dynamics \cite{bio2} and so on. \par 
Despite their popularity, the ability of the sine-Gordon and $\phi^4$ potentials to provide a fair description of physical processes for which they were proposed has been questioned in a great number of physical contexts. In crystal lattices for instance, structural transitions occur in an environment where characteristic features of the lattices are subjected to frequent changes. These changes are caused by constraints that are manifest in variations of the equilibrium configurations of atoms/molecules along the crystal lattice, or of their characteristic energies at equilibrium (i.e. the energy barrier) and so on. Some concrete physical situations are the variations in hydrostatic pressures (due for instance to isotopic substitutions or chemical reactions), change in temperature of the system, etc. that induce changes in bond lengths (bond stretching or compressions), in the activation energy in barrier-crossing processes (lowering or increase of the potential barrier), and so on \cite{cros1,cros2,crosb,cros3,cros3a,cros4}. To account for the later processes, improved variants of the sine-Gordon and $\phi^4$ potentials with tunable shape profiles have been proposed. Thus, Remoissenet and Peyrard introduced a familty of periodic potentials (the well-known Remoissenet-Peyrard potentials) \cite{remoi}, whose characteristic features such as the period, the potential amplitude and the steepness of the potential walls, can be controlled by a deformability parameter. Concerning the $\phi^4$ potential, parametrized double-well (DW) potentials have been proposed including polynomial potentials \cite{mag1,schm} and hyperbolic DW potentials \cite{kon,kon1,raz}. While these parametrized DW potentials have a deformable shape profiles, their link with the standard $\phi^4$ model is not always easy to establish. To effectively deal with a parametrized DW potential that can be reduced smoothly to the $\phi^4$ model, in some previous works we proposed a family of hyperbolic potentials whose DW shapes could be tuned distinctively, admitting the canonical $\phi^4$ potential as a specific limit \cite{dik1,dik2,dik3,dik4}. Peculiar features of some of these parametrized DW potentials (hereafter referred to as Dikand\'e-Kofan\'e double-well (DKDW) potentials) have been pointed out in recent studies \cite{zh1,zh2,zh3,zh4}. \par
In ref. \cite{bak3}, Jensen et al. established that although the $\phi^4$ model on a lattice is not integrable, the associate continuum periodic-kink solution could be generic of a chaos-free $2D$ discrete map with the Schmidt substrate potential \cite{schm}. Later on this result was extended \cite{zh5,zh6} to two members of the family of DKDW potentials, leading to two chaos-free maps \cite{zh5,zh6} that turned out to be two parametrized forms of the Schmidt map \cite{schm} and remarkably, belong to the well-known family of Quispel-Roberts-Thompson maps \cite{quisp} characterized by discreteness, low dimensionality and complete integrability \cite{quisp2}. Yet, even if members of the family of DKDW models on a lattice are non intgrable in the sense of Quispel–Roberts–Thompson integrability criteria \cite{quisp,quisp2,quisp3}, the fact that the models are parametrized introduces the perspective of a possible control of the occurance of unstable and chaotic orbits in the associate phase space. In particular the deformability parameter could be chosen in such a way that the map is always chaos free, similar with the recent study \cite{zh4} in which we established that exact eigenstates (eigenfunctions and corresponding energy eigenvalues) of the transfer operator associated with the statistical mechanics of one of the DKDW models, could be found for specific values of the deformability parameter. In the present context, the possibility to control the occurance of unstable trajectories and chaotic orbits would imply a non-universal character of route to chaos in the discrete DKDW maps, non universal as opposed to the universal picture related to the discrete $\phi^4$ map \cite{bak2}. The non-universality of the route to chaos will mean a Feigenbaum sequence \cite{bak1} and the structure of the bifurcation diagram dependent on the deformability parameter. \par
Our objective in the present work is to explore the phase-space texture and the structure of bifurcation diagrams for $2D$ discrete bistable maps, considering two different parametrized DKDW potentials. One has fixed minima but a variable barrier height, and the other has fixed barrier height but variable minima. We examine analytically the first bifurcations around the trivial fixed point, and establish conditions (where applicable) for a controllable route to chaos. While the first bifurcation point is continuously shifted to larger values of the ratio of the $\phi^4$-potential amplitude to the spring constant $K$ in the case of the discrete map with variable barrier and fixed minima, we find that for the second map the first pitchfork instability occurs at the same value of the potential amplitude obtained \cite{bak1} in the $\phi^4$ context. However, the area separating the first pitchfork bifurcation and the cascade of bifurcations marking the transition from period-two to period-four orbits, is gradually shrinked with increasing values of the deformability parameter. A global analysis of the two maps in phase space and with the help of bifurcation diagrams, will be carried out numerically for selected values of the deformability parameter.  

\section{\label{sec:level2}Discrete Hamiltonian and the family of $2D$ parametric maps} 
Consider a $1D$ periodic lattice of $N$ identical oscillators (which can be atoms, molecules, macromolecules, cells, etc.) of mass $m$, interacting them and others via identical linear springs of a spring constant $K$. The equilibrium configuration of the system under the linear two-body interactions, sketched in fig. \ref{figa}, is one in which oscillators are equally separated (hereafter the lattice spacing will be set to unity). In addition to the two-body interactions each oscillator is assumed to experience a one-body force, generated by an on-site potential created by the specific background substrate the physical system. The substrate potential will tend to keep individual oscillators at their equilibrium positions on the lattice sites $n$, thus opposing to their displacements due to two-body forces (springs here can be interatomic or intermolecular bonds, etc.). \par
Let $u_n$ be the displacement of the $n^{th}$ oscillator relative to its equilibrium position at the bottom of one of the two energetically degenerate potential wells. For such discrete systems, the total Hamiltonian can be expressed:
\begin{equation}
\label{hamil1}
H= \sum_{n=1}^{N}{\left[\frac{m}{2}\, \left(\frac{du_n}{dt}\right)^2 + \frac{K}{2}\left(u_{n+1} - u_n\right)^2 + V_{\mu}(u_n)\right]},
\end{equation}
where $t$ is time variable and $V_{\mu}(u_n)$ is the substrate potential. Here we are interested in a substrate potential with a double-well shape, corresponding to physical systems with bistable energy landscapes. The most familiar of such potential energy is the $\phi^4$ potential, its canonical form is (see e.g. \cite{remoi,sin2,krum}):
\begin{equation}
V(u)=\frac{a}{4}\left(u^2-1\right)^2, \label{f4}
\end{equation}
where the potential amplitude $a$ is real and constant. The $\phi^4$ field in eq. (\ref{f4}) possesses two degenerate minima $u_{1,2}=\pm 1$, and a maximum $u_0=0$ where the potential $V(0)=a/4$. As its characteristic parameters (i.e. its extrema and its maximum value) indicate, the $\phi^4$ field has two fixed minima and an extremum for which the value of the potential depends only on the real parameter $a$. In the context of second-order phase transition $a$ depends on temperature and is expected to vanish at the critical point. However, in real physical contexts the parameter $a$ will not always depend only on temperature. Indeed in most real contexts of atomic or molecular crystals, the equilibrium positions of atoms or molecules will constantly change as a result of variations of bond lengths and strengths, and hence of characteristic bond energies. Such changes are induced by a variation of hydrostatic pressures in the crystals, or interactomic/intermolecular pressures caused by isotopic substitutions, the presence of chemical catalyzers or chemical reactions occuring concomitantly with the structural transition \cite{pa}. Ideal systems in which such processes are expected include hydrogen-bonded atomic and molecular crystals prone to ferroelectric and antiferroelectric phase transitions \cite{p1,p2,p3,p4}. To take into consideration the possible variations of the potential amplitude and positions of the two potential minima, a family of parametrized double-well potentials was proposed in refs. \cite{dik1,dik2,dik3,dik4}. In one version the two degenerate minima can be tuned by varying a deformability parameter without changing the height of the potential barrier \cite{dik1}, in the other version \cite{dik2} the barrier height can be changed continuously without changing positions of the two degenerate minima, and in the third version both the dgenerate minima and the barrier height can change simultaneously \cite{dik4}. This family of parametrized double-well potential will be referred to as Dikand\'e-Kofan\'e potentials. The DKDW potential with a variable barrier height but fixed minima, was considered recently by Bazeia et al \cite{zh2} in their study of oscillons induced by kink-antikink scatterings in double-well system with complex double-well shape profiles. In ref. \cite{zh1}, Zhou et al. established that quantum Hamiltonians with the DKDW potential with fixed minima and variable potential barrier, provide an ideal model for bistable systems that can display simultaneously first-order and second-order transitions in quantum tunnellings. \par
\begin{figure*}
\begin{minipage}{0.5\textwidth}
\includegraphics[width=2.8in,height=2.4in]{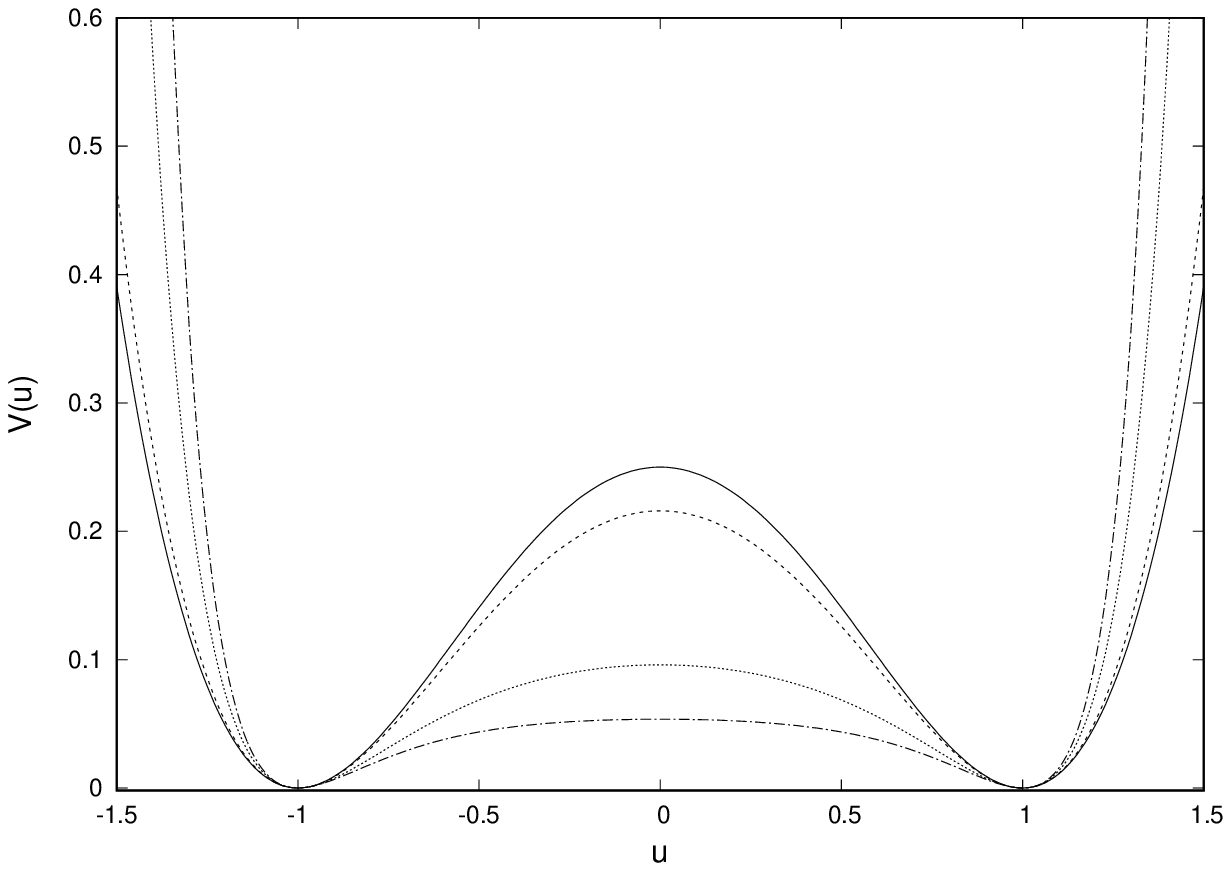}
\end{minipage}%
\begin{minipage}{0.5\textwidth}
\includegraphics[width=2.8in,height=2.4in]{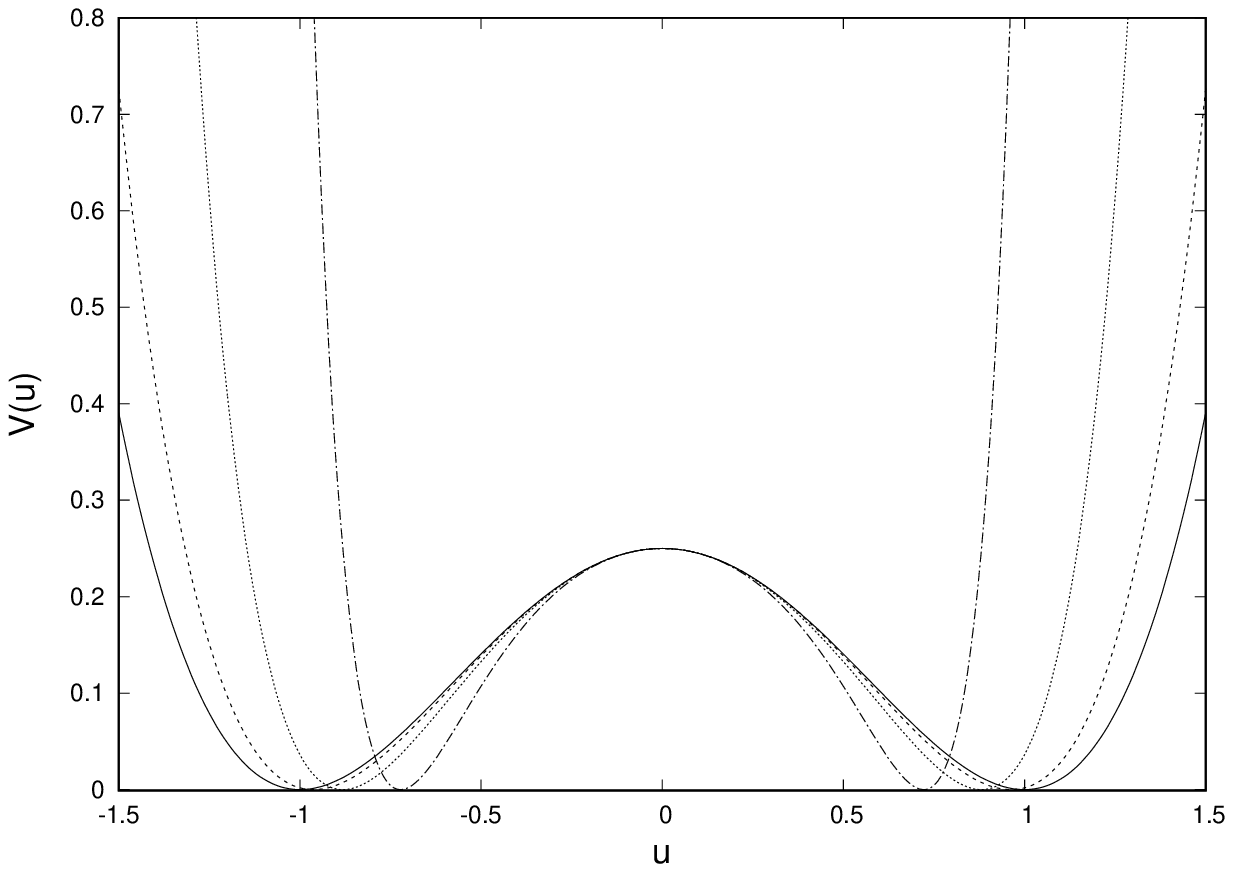}
\end{minipage}
\caption{\label{fig1}(Color online) Parametrized double-well potential $V_{\mu}(u)$ versus $u$, for different values of $\mu$. Left panel: DKDW potential with variable barrier and fixed minima (from highest to lowest potential barriers, $\mu=0$, 0.5, 2, 4). Right panel: DKDW potential with variable minima and fixed barrier height (from the minima nearest to $u=0$ to the farthest, $\mu=4$, 2, 0.5, 0).} 
 \end{figure*}

 
In this work we shall focus on two particular members of the family of DKDW potentials, i.e. the one with variable barrier height but fixed minima and the one with variable minima but fixed barrier height. Starting, let us write down the general expression of the family of DKDW potentials \cite{dik4}:
\begin{equation}
V_{\mu}(u)= \frac{a_{\mu}}{4} \left[\frac{1}{\mu^2} \sinh^2\left(\alpha_{\mu} u\right) - 1\right]^2, \label{eq1} 
\end{equation}
where $a_{\mu}$ and $\alpha_{\mu}$ are two real functions. For the two members we have chosen to study, these functions are defined as:
\begin{itemize}
 \item DKDW potential with variable potential barrier but fixed positions of the two degenerate minima \cite{dik1}:
 \begin{equation}
a_{\mu}=a\,\frac{\mu^2}{(1+\mu^2)\,arsinh^2(\mu)}, \hskip 0.3truecm \alpha_{\mu}= arsinh(\mu). \label{pt1} 
\end{equation}
 \item DKDW potential with fixed barrier but variable positions of the two degenerate minima \cite{dik2}:
 \begin{equation}
a_{\mu}=a, \hskip 0.3truecm \alpha_{\mu}= \mu. \label{pt2} 
\end{equation}
\end{itemize}
In these functions the quantity $\mu$ is the deformability paremeter, it is real and defined such that eq. (\ref{eq1}) reduces to the $\phi^4$ potential (\ref{f4}) when $\mu\rightarrow 0$. The two different DKDW potentials $V_{\mu}(u)$ given in eqs. (\ref{pt1}) and (\ref{pt2}), are plotted in fig. \ref{fig1}, for some arbitrary values of the deformability parameter $\mu$ listed in the caption. The left graph shows the DKDW potential with variable barrier height but fixed positions of the potential minima i.e. $u_{1,2}=\pm 1$, while the right graph represents the DKDW potential with fixed barrier height but variable positions of the two degenerate minima given by:
\begin{equation}
u_{1,2}=\pm \frac{arsinh(\mu)}{\mu}. \label{minb}
\end{equation}

As our main interest is the spatial configurations of the system of N oscillators favored by the competition between the two-body and one-body forces, we shall ignore the kinetic energy term in total Hamiltonian. With this consideration the equilibrium configurations of the system will be determined by minimizing the discrete Hamiltonian eq. (\ref{hamil1}) with respect to $u_n$, which yields:
\begin{equation}
\label{eq2}
u_{n+1} -2 u_n + u_{n-1} = \frac{1}{K}\frac{\partial{V_{\mu}(u_n)}}{\partial{u_n}}, 
\end{equation}
which is a nonlinear difference equation. Instructively, this nonlinear difference equation stands in principle as a generalization of the $\phi^4$ model and hence is not expected to be integrable, even in the continuum limit. Note that in this last limit the nonlinear difference equation (\ref{eq2}) possesses \cite{zh1,zh6} a periodic-kink solitary-wave solution. However, to efficiently address the issue of integrability of the current problem in the presence of deformability, it is more enriching to resort to an analysis of the texture of trajectories in phase space as well as of the structure of bifurcation diagrams characterizing routes to chaos in the system. To this end we transform the nonlinear difference equation (\ref{eq2}) into a $2D$ discrete map of the form:
\begin{eqnarray}
{\it \mathcal{T}_{\mu}}: \hskip 0.2truecm u_{n+1} &=& 2 u_n - v_n + \frac{1}{K}\frac{\partial{V_{\mu}(u_n)}}{\partial{u_n}}, \nonumber \\ v_{n+1}&=& u_n. \label{eq3}
\end{eqnarray}
In the language of dynamical system theory, the last system represents a $2D$ discrete map with the DKDW potential. Since the underlying nonlinear function is parametrized, we shall refer to the $2D$ system as a parametric map. Note that when $\mu\rightarrow 0$, the discrete parametric map reduces to the map studied by Bak and Jensen in ref. \cite{bak1}, and by Bak and Prokovsky in ref. \cite{bak2} in the context of metal-insulator transition in half-filled Peierls systems. In the next section we examine fixed points of the map including their stability, and the influence of shape deformability of the potentials on the two first bifurcations for the two maps. Although the parametrization complexifies the problem and hence restricts the ability for analytical formulation of aspects such as the generation of Feigenbaum number sequences associated with successive period-doubling bifurcations \cite{bak1}, we will be satisfied with the analytical formulation of the parametric dependence of characteristic values of the ratio $a/K$ forming the Feigenbaum sequences. This parametric dependence clearly indicates a non-universal character for these sequences in the particular case of the two parametric maps studied in this work.   

\section{\label{three}Stability of fixed points and the two first sequences of bifurcations}
The transformation $\mathcal{T}_{\mu}$ maps a point in the phase space $(u, v)$, onto another point in the same phase space. The map $\mathcal{T}_{\mu}$ possesses a fixed point at $(u,v)=(0, 0)$ where the substrate potential $V_{\mu}(u_n)$ is maximum whatever the value of the deformability parameter $\mu$, we call it a trivial fixed point. In addition there are two fixed points at $(u,v)=\pm (1,1)$, corresponding to minima of the parametrized substrate potential. In the previous study dealing with the $\phi^4$ map \cite{bak1}, it was pointed out that the most interesting and richest dynamical processses associated with bistable systems were those associated with the fixed point $(0,0)$. In this previous study it was shown that the $\phi^4$ map was area preserving, and that for $a=4K$ the fixed point $(0,0)$ undergoes a transition from an elliptical fixed point to an hyperbolic fixed point, via a pitchfork bifurcation. At this pitchfork instability, simple-periodic trajectories in phase space become unstable given birth to period-two orbits involving two new fixed points. These new fixed points are determined by two transformations generating the following equation:
\begin{eqnarray}
(u,v)&=& \mathcal{T}_{0}\left(\mathcal{M}_{0}(u,v)\right) \nonumber \\
&=&\mathcal{T}^2_{0}(u,v), \label{pf2} 
\end{eqnarray}
which for the $\phi^4$ map admits two pairs of solutions namely \cite{bak1}:
\begin{equation}
(u,v)=\pm\left(\sqrt{1-\frac{4}{\tilde{a}}}, -\sqrt{1-\frac{4}{\tilde{a}}}\right), \hskip 0.25truecm \tilde{a}=a/K. \label{rap}  
\end{equation}
Consider the same problem for the $2D$ parametric map given by eq. (\ref{eq3}), assuming the fixed point $(0,0)$. In this goal we start by observing that the first bifurcation associated with this fixed point is best formulated by linearizing eq. (\ref{eq3}) around $(0,0)$, and analyzing the set of tangent space orbits \cite{greene1} $(\delta u, \delta v)$ around the fixed point generated by the $2\times 2$ linear matrix equation:
\begin{equation}
\left(\delta u_{n+1}, \delta v_{n+1}\right)=\mathcal{M}_{\mu}\left(\delta u_n, \delta v_n\right), \label{mapa} 
\end{equation}
where: 
\begin{equation}
\mathcal{M}_{\mu} = \left(
                    \begin{array}{cc}
                        2+\frac{1}{K}\frac{\partial^2{V_{\mu}(u_n)}}{\partial{u^2_n}} & -1\\
                        1 & 0

                        \end{array}
                        \right) \label{a}
\end{equation}
is the $2\times 2$ transformation matrix. This matrix possesses two eigenvalues which can be expressed most generally, in terms its trace $\mathrm{Tr}\left[\mathcal{M}_{\mu}\right]$ and its Jacobian $\det\left[\mathcal{M}_{\mu}\right]$ as follows:
\begin{equation}
\lambda_{1,2}=\frac{\mathrm{Tr}\left[\mathcal{M}_{\mu}\right] \pm\sqrt{\left(\mathrm{Tr}\left[\mathcal{M}_{\mu}\right]\right)^2 - 4\det\left[\mathcal{M}_{\mu}\right]}}{2}, \label{eigvs}
\end{equation}
According to Greene \cite{greene1} any $2D$ discrete mapping for which $\det(\mathcal{M}_{\mu}) \leq 1$, is area preserving. This is the case for the $2\times 2$ matrix eq. (\ref{a}), for which $\det (\mathcal{M}_{\mu}) = 1$. It follows that eigenvalues of this matrix will depend only on its trace, i.e. the pair of eigenvalues given by eq. (\ref{eigvs}) simplifies to: 
\begin{equation}
\lambda_{1,2}=\frac{\mathrm{Tr}\left[\mathcal{M}_{\mu}\right] \pm\sqrt{\left(\mathrm{Tr}\left[\mathcal{M}_{\mu}\right]\right)^2 - 4}}{2}, \label{eigv}
\end{equation}
where the trace:
\begin{equation}
\mathrm{Tr}\left[\mathcal{M}_{\mu}\right]=  2+\frac{1}{K}\frac{\partial^2{V_{\mu}(u)}}{\partial{u^2}}\vert_{u=0}. \label{trs}
\end{equation}
 Learning from formula (\ref{eigv}), it follows that period-one orbits around the fixed point $(0,0)$ will be stable provided the condition: 
\begin{equation}
 \left(\mathrm{Tr}\left[\mathcal{M}_{\mu}\right]\right)^2 - 4\leq 0. \label{fpa}
 \end{equation}
 For the discrete DKDW map with variable barrier height and fixed minima, the stability condition (\ref{fpa}) implies that the first bifurcation will occur for values of $\tilde{a}$ determined by the relation:
 \begin{equation}
\mu^2 = \frac{\tilde{a}}{4}-1, \label{fia} 
\end{equation}
with $a$ and $\mu$ positives. For the DKDW map with fixed barrier height and variable minima, the stability condition (\ref{fpa}) suggests that the first bifucation will occur at $\tilde{a}=4$ as with the $\phi^4$ map \cite{bak1}. \par
\begin{figure*}\centering
\begin{minipage}{0.49\textwidth}
\includegraphics[width=2.7in,height=2.2in]{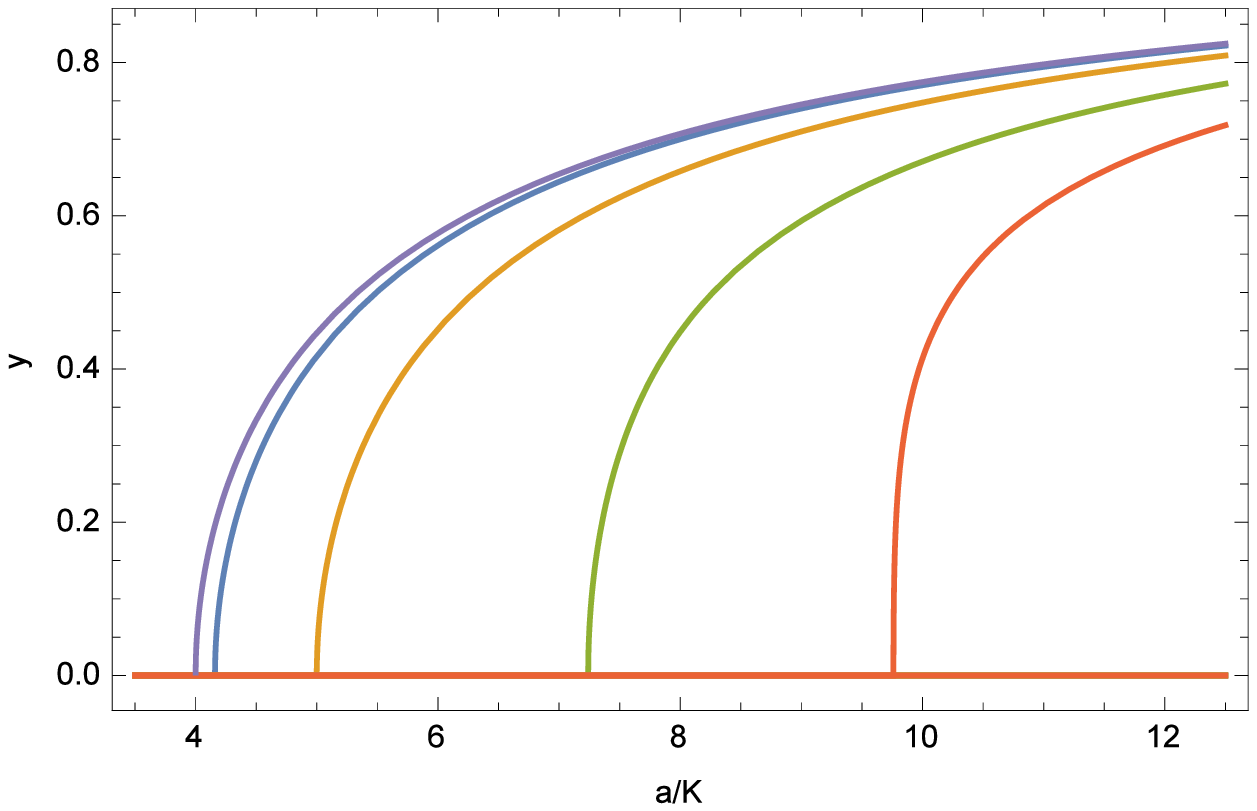}
\end{minipage}%
\begin{minipage}{0.49\textwidth}
\includegraphics[width=2.7in,height=2.2in]{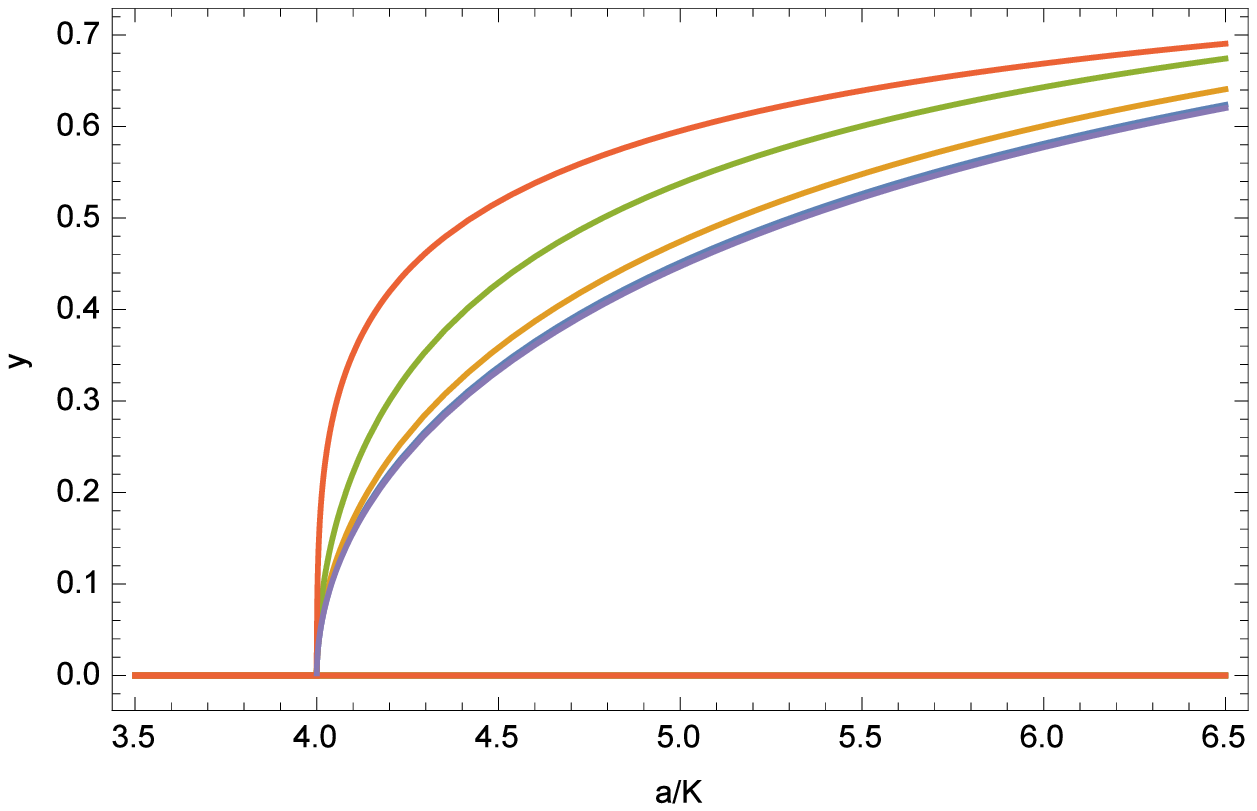}
\end{minipage}
\caption{\label{figc}(Color online) Variation of $y$ extracted numerically from eq. (\ref{rot}), as a function of the ratio $a/K$ for different values of $\mu$. Left panel corresponds to the DKDW potential with variable barrier height and fixed minima, and the left panel corresponds to the DKDW potential with variable minima but fixed barrier height. Left graph, from left to right curves: $\mu=0$, 0.2, 0.5, 0.9, 1.2.  Right graph, from top to bottom curves: $\mu=0$, 0.2, 0.5, 0.9, 1.2.} 
 \end{figure*}

When starting this section we recalled that for the $\phi^4$ map, the two fixed points of period-two orbits generated from instability of period-one orbits around the fixed point $(0,0)$, were obtained from two iterations (i.e. eq. (\ref{pf2})) and were given by the two pairs of points obtained analytically in eq. (\ref{rap}). Finding similar analytical expressions for the two parametric maps under study is not an easy task, nevertheless we can exploit symmetry properties of the map, which reveal that these two new fixed points, that we denote $(u_0,v_0)$, obey $(u,v)=\pm (y, -y)$ where $y$ is a real and positive function of $\tilde{a}$ and $\mu$. With this assumption, the period-two fixed-point equation (\ref{pf2}) leads to the following transcendental equation;   
\begin{equation}
2y + \frac{1}{K}\frac{\partial{V_{\mu}(y)}}{\partial{y}}=0, \label{rot}
\end{equation}
the roots of which determine $y$. To extract these roots we make use of the Brent algorithm \cite{brent}, a robust root-finding rule that combines the bracketing, bisection and inverse quadratic interpolation methods \cite{brent}. Brent's scheme turns out to be very efficient when seeking for roots of non-polynomial functions, which may display multiple dsingularities. Fig. \ref{figc} show the numerical solution to eq. (\ref{rot}), for the parametric map with the DKDW potentials (\ref{eq1})-(\ref{pt1}) (left graph) and (\ref{eq1})-(\ref{pt2}) (right graph).   

 More explicitly the two graphs of fig. \ref{figc} represent $y$ as a function of $\tilde{a}$, for $\mu=0$, 0.5, 1, 1.5. The left graph shows that when $\mu=0$, the variation of $y$ with the ratio $\tilde{a}$ reproduces the analytical expression $y=\sqrt{1-\frac{4}{\tilde{a}}}$ present in eq. (\ref{rap}). As the deformability parameter $\mu$ is increased, fixed points of period-two orbits are gradually shifted forward. This is also the case for values of $\tilde{a}$ for which the first bifurcation is expected, indeed eq. (\ref{fia}) clearly indicates that values of $\tilde{a}$ at the first bifurcation will be increased as we increase $\mu$. On the other hand, the right graph of fig. \ref{figc} suggests that fixed points of period-two orbits for this map will always vanish at $\tilde{a}=4$ whatever the value of $\mu$. However when $\tilde{a}>4$, increasing $\mu$ contracts the region between the instability points of period-one orbits (i.e. $\tilde{a}=4$ for any $\mu$) and period-two orbits. 
\section{\label{sec} phase-space texture and structures of bifurcation diagrams}
In the previous section we discussed characteristic features of the fixed point $(0,0)$ for the two $2D$ discrete parametric maps, as well its stability and the influence of shape deformability of the double-well potentials on the two first bifurcations of the corresponding parametric maps. However, a global evolution of the two maps is required to gain a full undertanding of the overall texture of the phase space, and the structure of bifurcation diagrams of these maps taking into account their distinct parametrizations. We first consider the phase-space dynamics of the two maps, starting from $\mu=0$ and increasing the ratio $\tilde{a}$ around its critical value $\tilde{a}_c=4$ corresponding to the first period-doubling bifurcation point of the $\phi^4$ map. \par
\begin{figure}\centering
\begin{minipage}{0.5\textwidth}
\includegraphics[width=2.8in,height=2.2in]{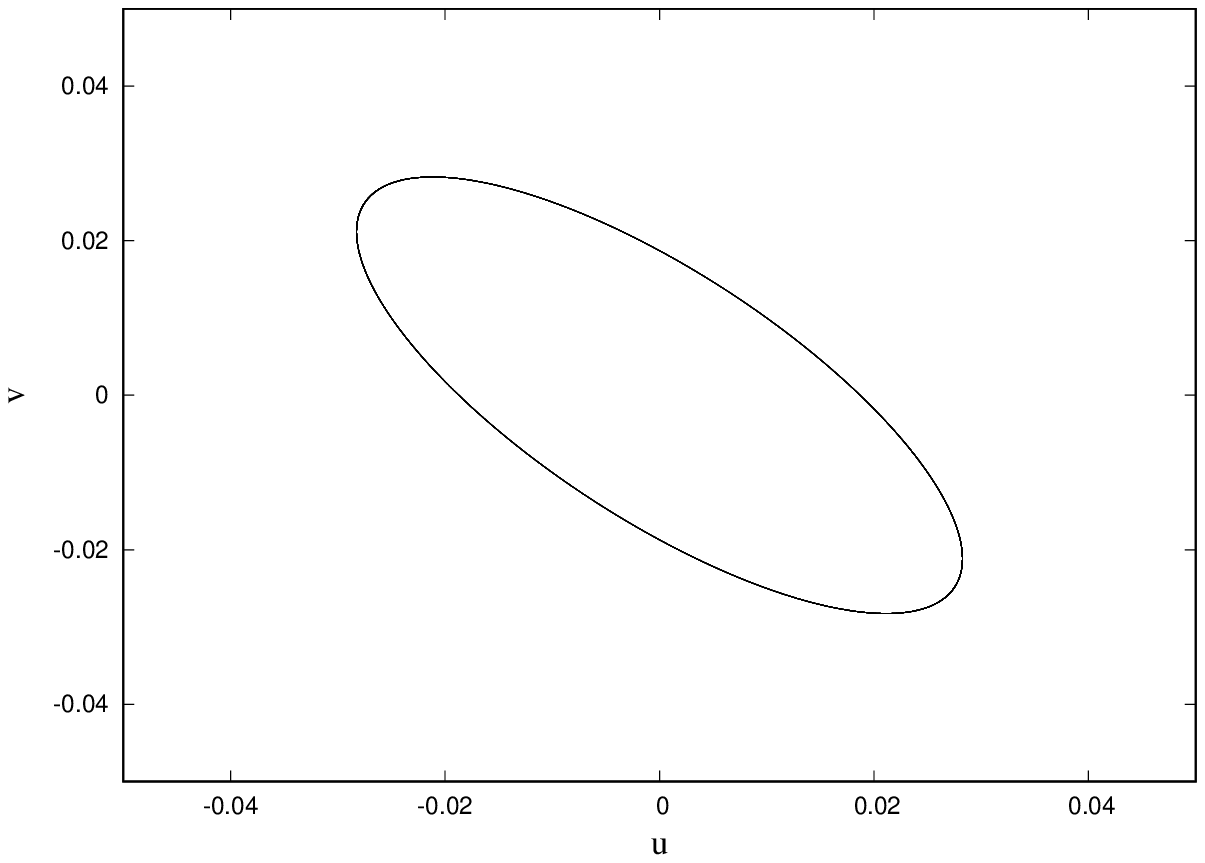}(a) 
\end{minipage}%
\begin{minipage}{0.5\textwidth}
\includegraphics[width=2.8in,height=2.2in]{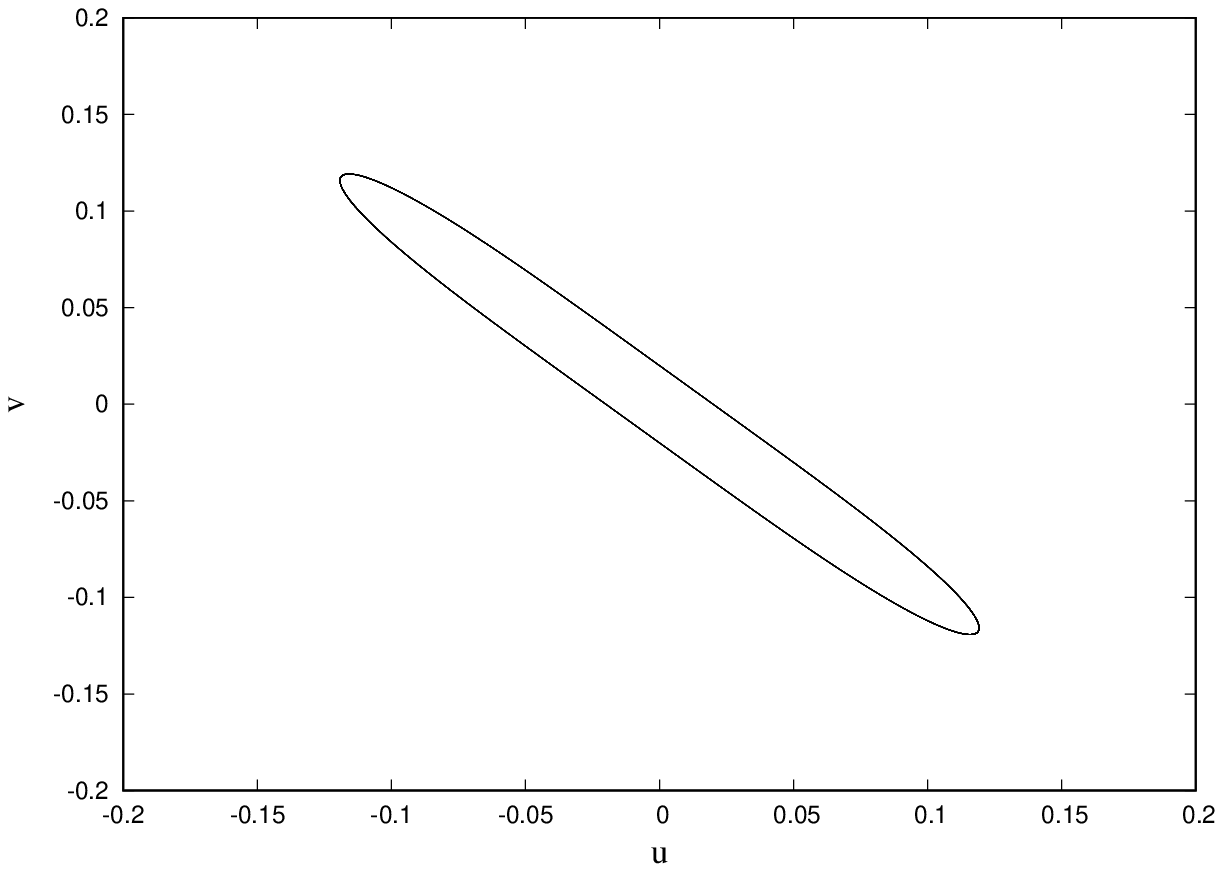}(b) 
\end{minipage}\vskip 0.8truecm
\begin{minipage}{0.5\textwidth}
\includegraphics[width=2.8in,height=2.2in]{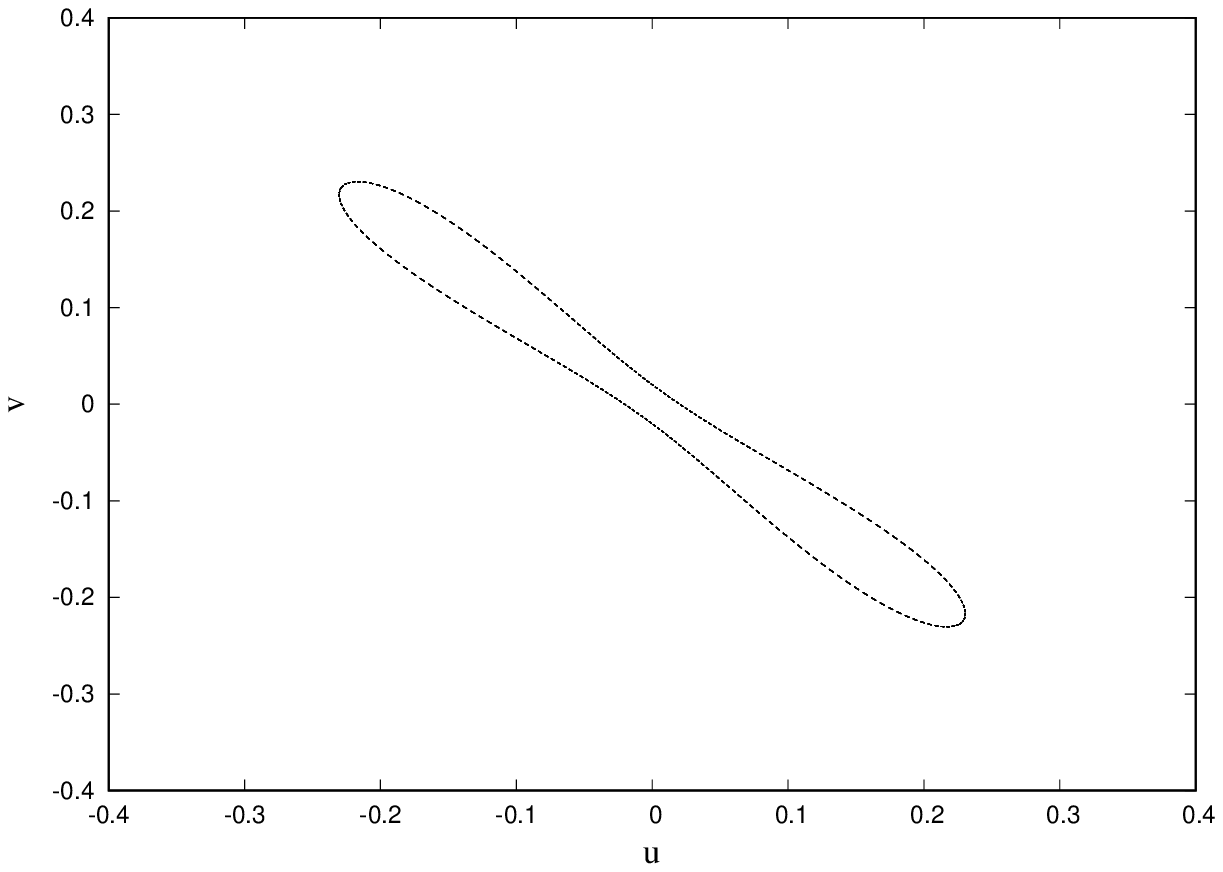}(c) 
\end{minipage}%
\begin{minipage}{0.5\textwidth}
\includegraphics[width=2.8in,height=2.2in]{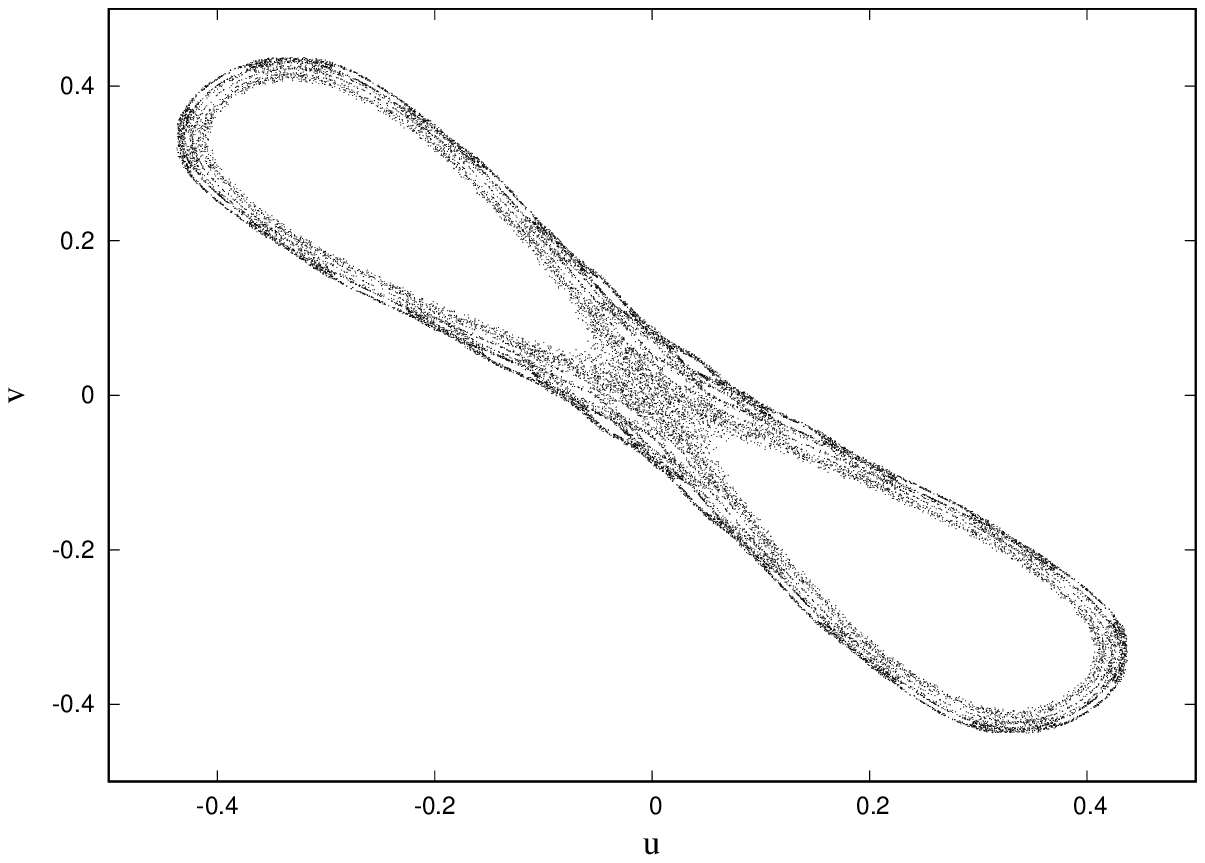}(d) 
\end{minipage}
\caption{\label{fpph1}(Color online) Trajectories in phase space for the $2D$ discrete parametric map with DKDW having a variable barrier height but fixed potential minima. Graph (a): Elliptic orbits ($\tilde{a}=3.5$), graph (b): unstable elliptic orbits ($\tilde{a}=3.9$), graph (c): nascent hyperbolic orbit ($\tilde{a}=4.1$), graph (d): hyperbolic orbits around two well separated fixed points ($\tilde{a}=4.35$). Here $\mu=0$ coinciding with the $\phi^4$ map.} 
 \end{figure}

 Fig. \ref{fpph1} represents trajectories around the fixed point $(0,0)$ for the DKDW map with variable barrier height but fixed mimina, when $\mu=0$ corresponding to the $\phi^4$ map. 
 Graph (a) features a stable elliptic fixed point ($\tilde{a}=3.5$), in graph (b) elliptical trajectories are nearly decaying ($\tilde{a}=3.9$), graph (c) shows a nascent hyperbolic orbit around two distinct fixed points ($\tilde{a}=4.1$), in graph (d) the fixed point $(0,0)$ has now given birth to two new elliptical fixed points ($\tilde{a}=4.35$). In the graph these two new fixed points are approximately at $\pm (0.2, -0.2)$, in agreement with the symmetry considerations that led to the transcendental equation in $y$ given (\ref{rot}), the solution of which was obtained numerically using Brent's algorithm. It is remarkable that in graph (d) of fig. \ref{fpph1}, the separatrix consists of dense orbits intersected by forbidden islands which enclose sparse, isolated orbits trapped inside the islands. Many other interesting motifs can be generated in the texture of phase space, as obtained in ref. \cite{bak1} for higher values of $\tilde{a}$. However this is not our objective here, let us rather see how the texture in fig. \ref{fpph1} is affected by the deformability of the double-well substrate. \par
\begin{figure*}\centering
\begin{minipage}{0.5\textwidth}
\includegraphics[width=2.8in,height=2.2in]{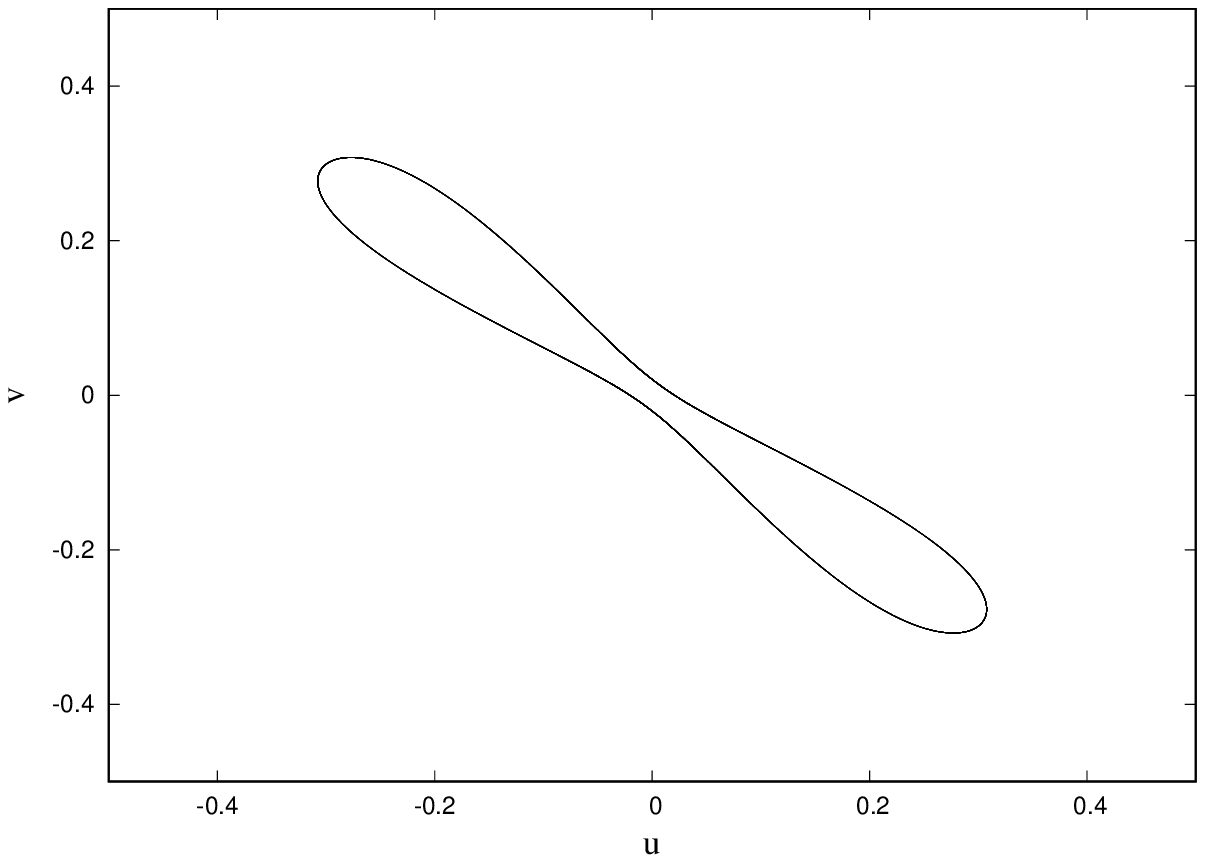}
\end{minipage}%
\begin{minipage}{0.5\textwidth}
\includegraphics[width=2.8in,height=2.2in]{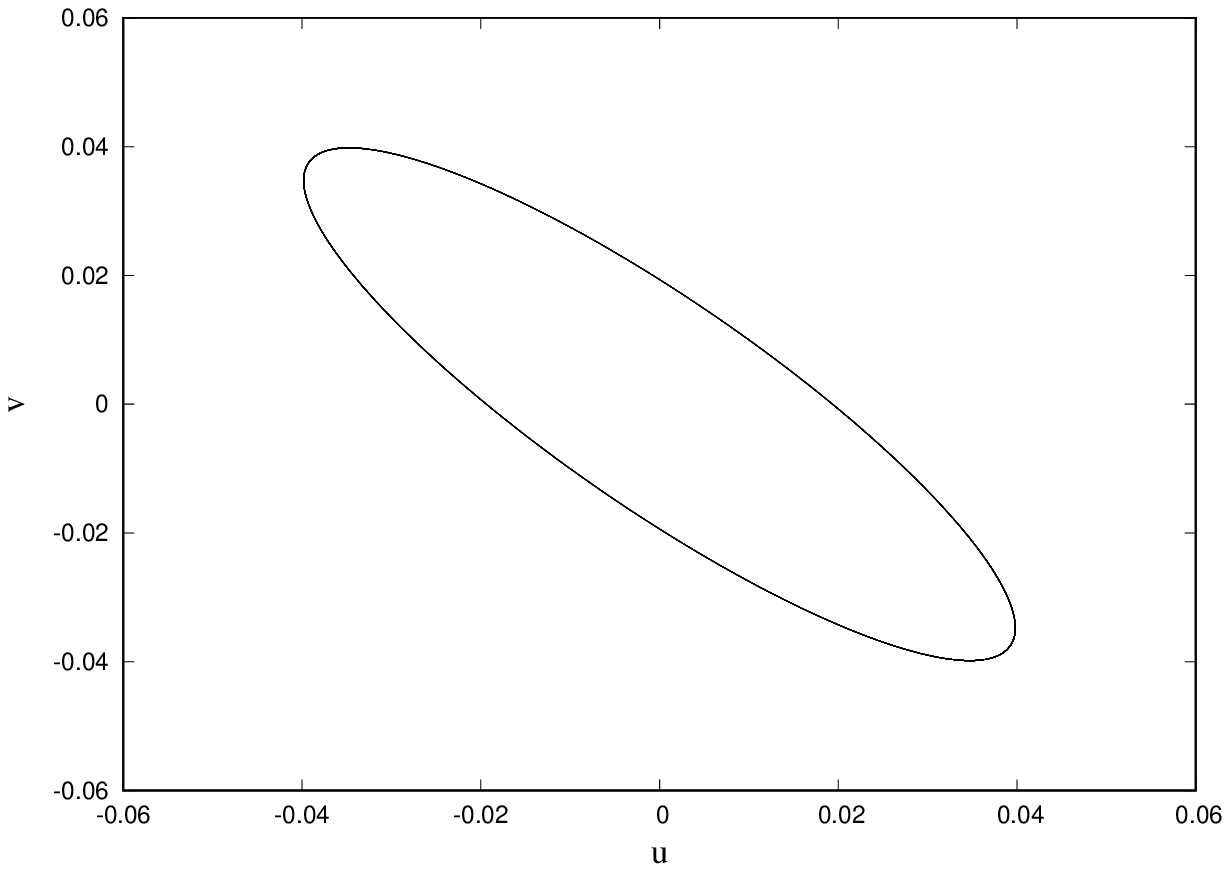}
\end{minipage}
\caption{\label{fpph2}(Color online) Replot of graph (d) in fig. \ref{fpph1}, now for $\mu=0.2$ (left graph) and $\mu=0.4$ (right graph). } 
 \end{figure*}

 To point out the effect of the deformability parameter on the phase space texture for this first map, in fig. \ref{fpph2} we reploted the graph \ref{fpph1}(d) for $\mu=0.2$ (left graph) and $\mu=0.4$ (right graph). One clearly sees that when $\mu=0.2$, the texture of phase space is reminiscent of the nascent hyperbolic orbit observed in graph fig. \ref{fpph1}(c) for $\mu=0$ and $\tilde{a}=4.1$. As the deformability parameter is increased, namely when $\mu=4$, the system is taken back to the phase-space texture dominated by elliptic orbits around the fixed point $(0,0)$.\par 
Turning to the second map (i.e. the DKDW map with variable mimima but fixed barrier height), the associated phase space texture will be similar to the one of the first map when $\mu=0$ (a value at which the two maps reduce together to the $\phi^4$ map). Therefore we shall examine only the influence of nonzero values of $\mu$ on the texture of phase space, focusing on the point $\tilde{a}=4$ and $\mu=0$ which is actually the graph in fig. {\ref{fpph1}(d). Thus, in fig. \ref{fpph3} we represent the phase space of the DKDW map with variable minima but fixed barrier height for $\tilde{a}=4$, with $\mu=0.2$ (left graph) and $\mu=0.4$ (right graph). We see that a variation of $\mu$ has relatively less effect on the phae space texture compared, at least with what we observed for the DKDW map with variable barrier height but fixed minima.  
\begin{figure*}\centering
\begin{minipage}{0.5\textwidth}
\includegraphics[width=2.8in,height=2.2in]{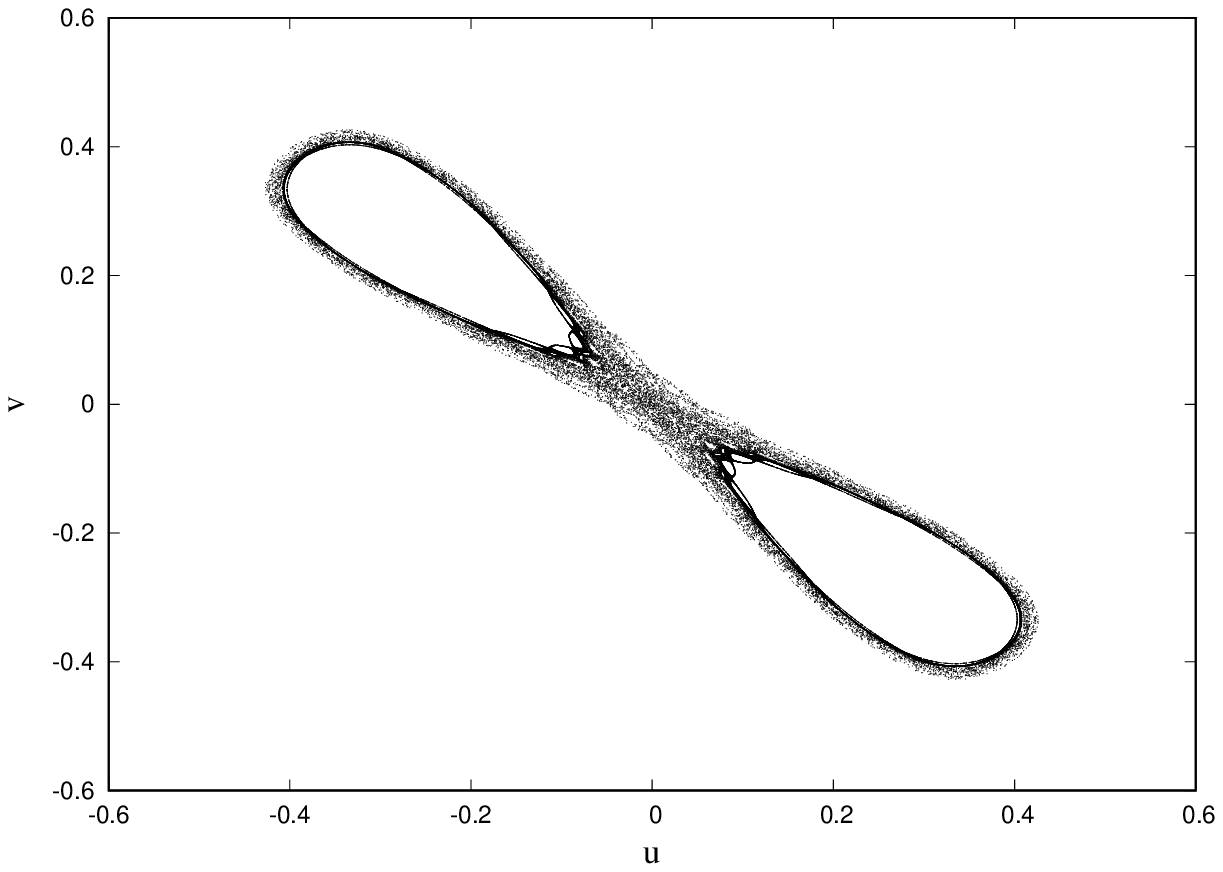}
\end{minipage}%
\begin{minipage}{0.5\textwidth}
\includegraphics[width=2.8in,height=2.2in]{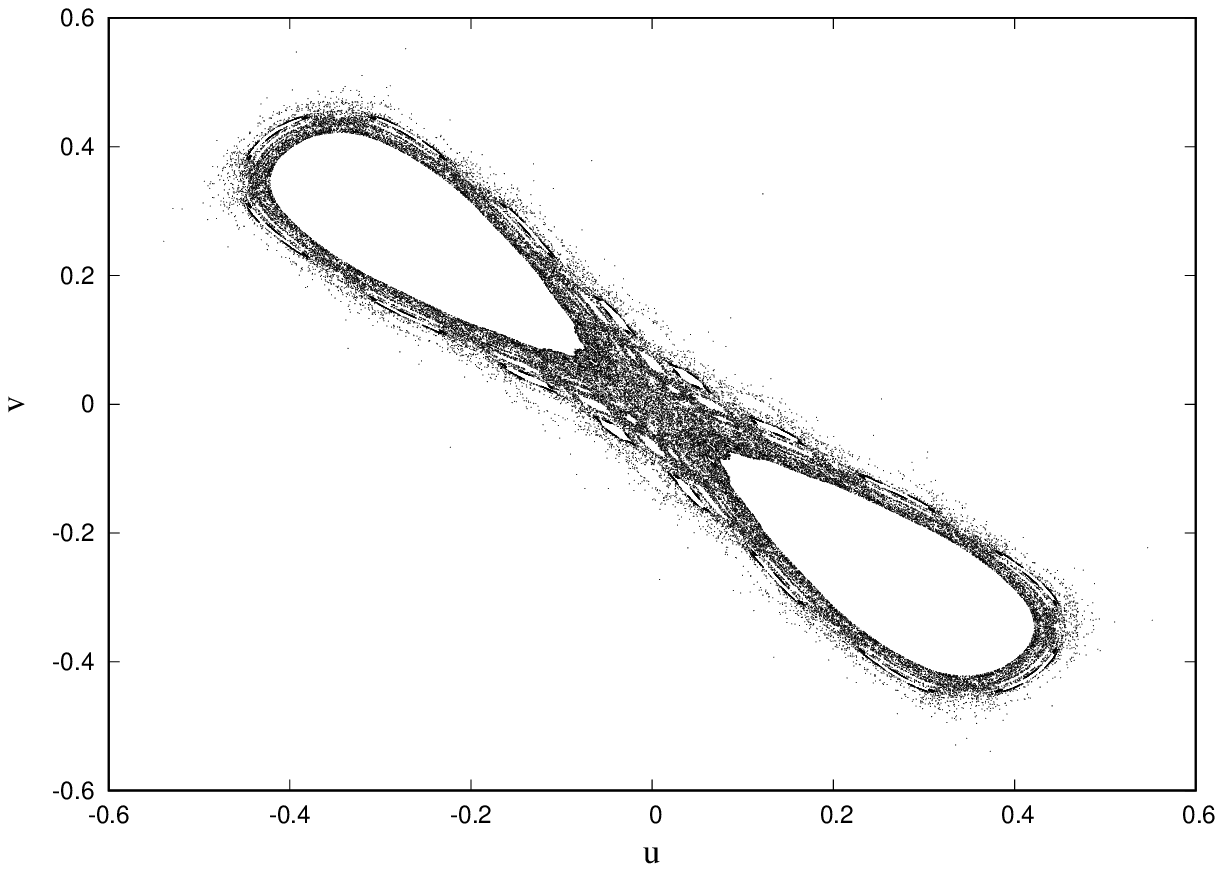}
\end{minipage}
\caption{\label{fpph3}(Color online) Texture of phase space for the DKDW map with variable minima and fixed barrier, for $\tilde{a}=1$ and $\mu=0.2$ (left graph), $\mu=0.4$ (right graph).} 
 \end{figure*}
\par The last aspect we shall examine is the structure of bifurcation diagrams of the two maps, with emphasis on their distinct parametrizations. Fig. \ref{fbifa} displays the bifurcation diagram with respect to $\tilde{a}$, of the DKDW map with variable barrier height for four different values of $\mu$ i.e. $\mu=0$ (a), $\mu=0.2$ (b), $\mu=0.5$ (c) and $\mu=1$ (d). The main insight from this series of graphs is the observed forward shift of values of $\tilde{a}$ at the first period-doubling bifurcation, as $\mu$ is increased. This behavior is of course consistent with the stability condition given analytically in formula (\ref{fia}).
\begin{figure}\centering
\begin{minipage}{0.5\textwidth}
\includegraphics[width=2.8in,height=2.2in]{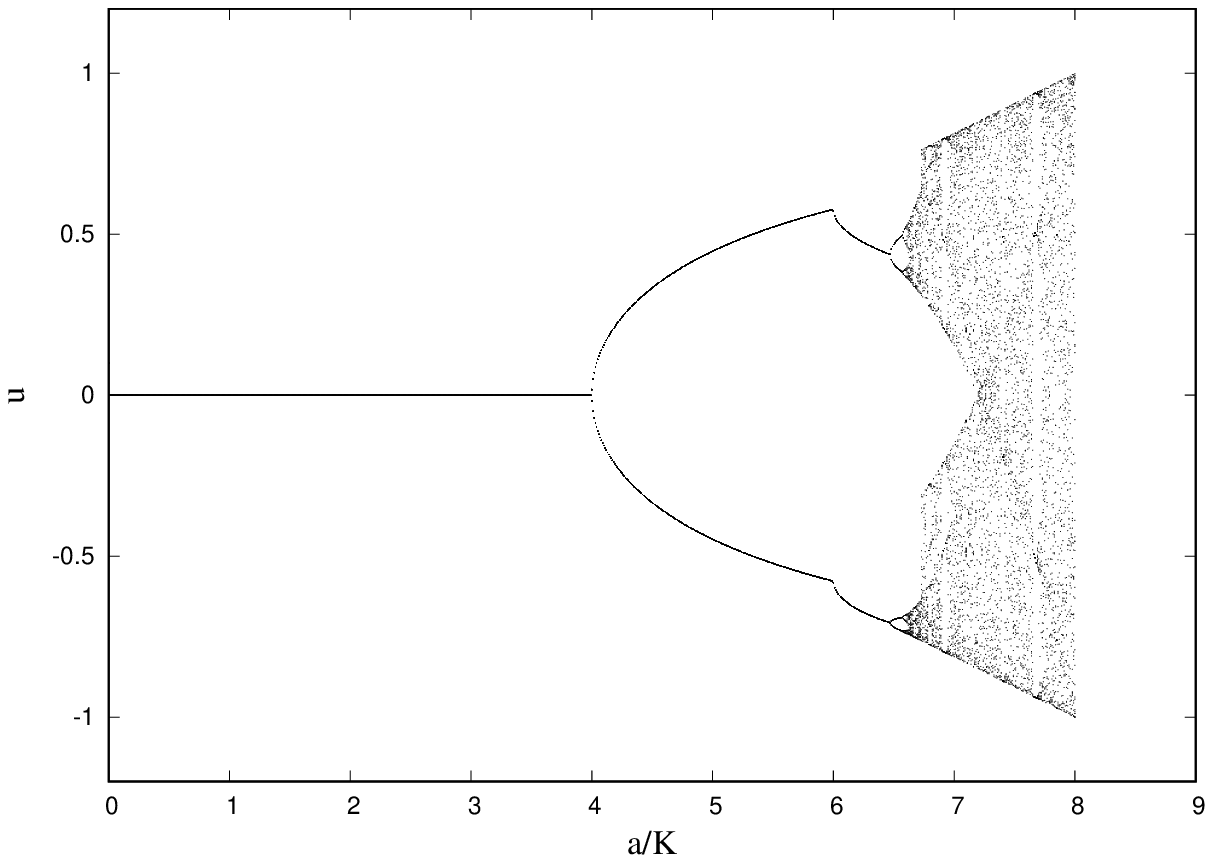}(a) 
\end{minipage}%
\begin{minipage}{0.5\textwidth}
\includegraphics[width=2.8in,height=2.2in]{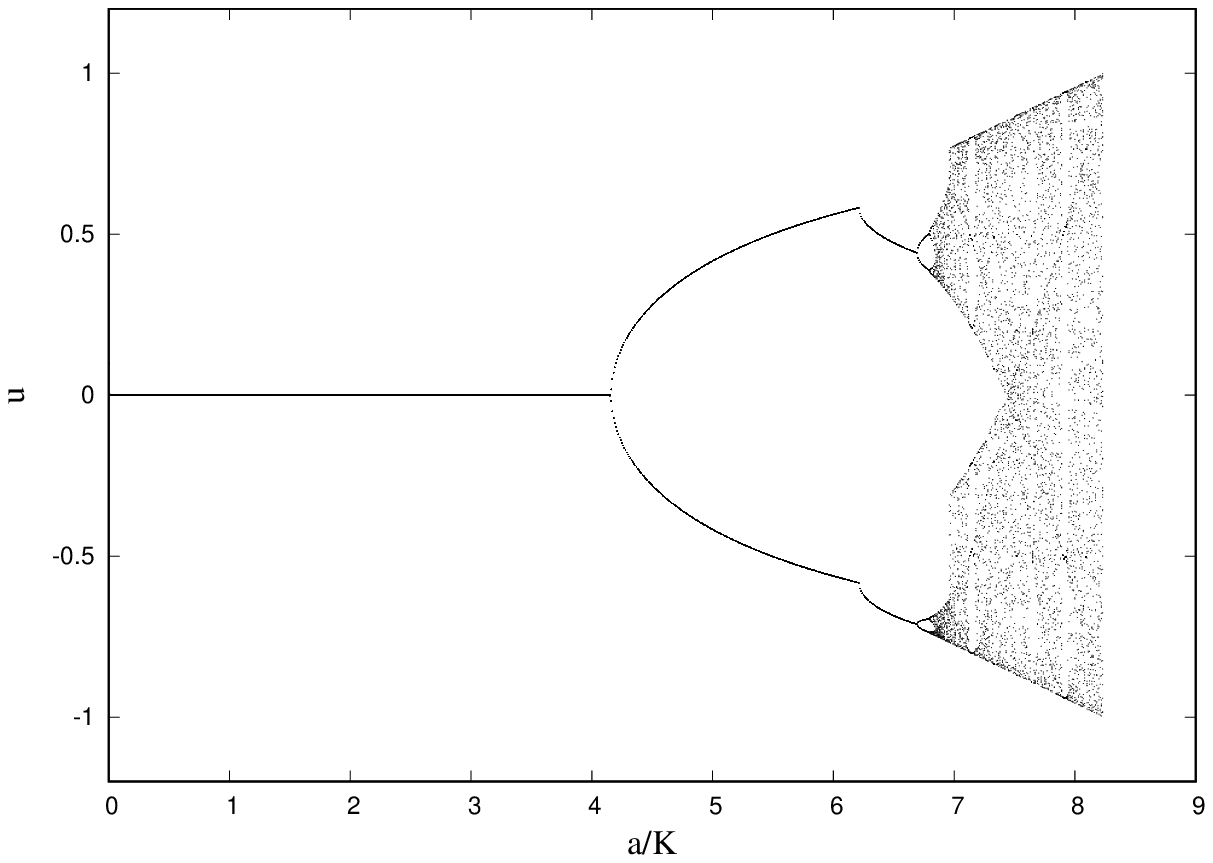}(b) 
\end{minipage}\vskip 0.8truecm
\begin{minipage}{0.5\textwidth}
\includegraphics[width=2.8in,height=2.2in]{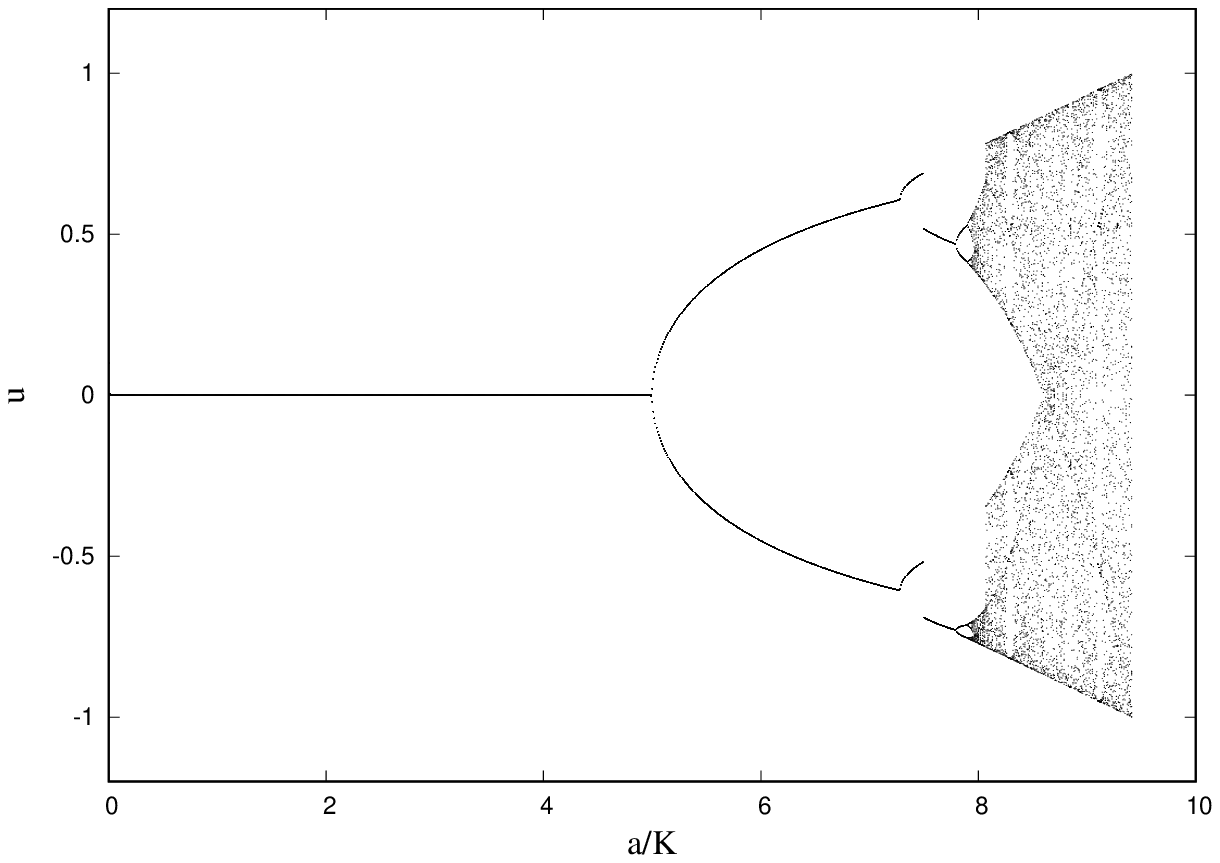}(c) 
\end{minipage}%
\begin{minipage}{0.5\textwidth}
\includegraphics[width=2.8in,height=2.2in]{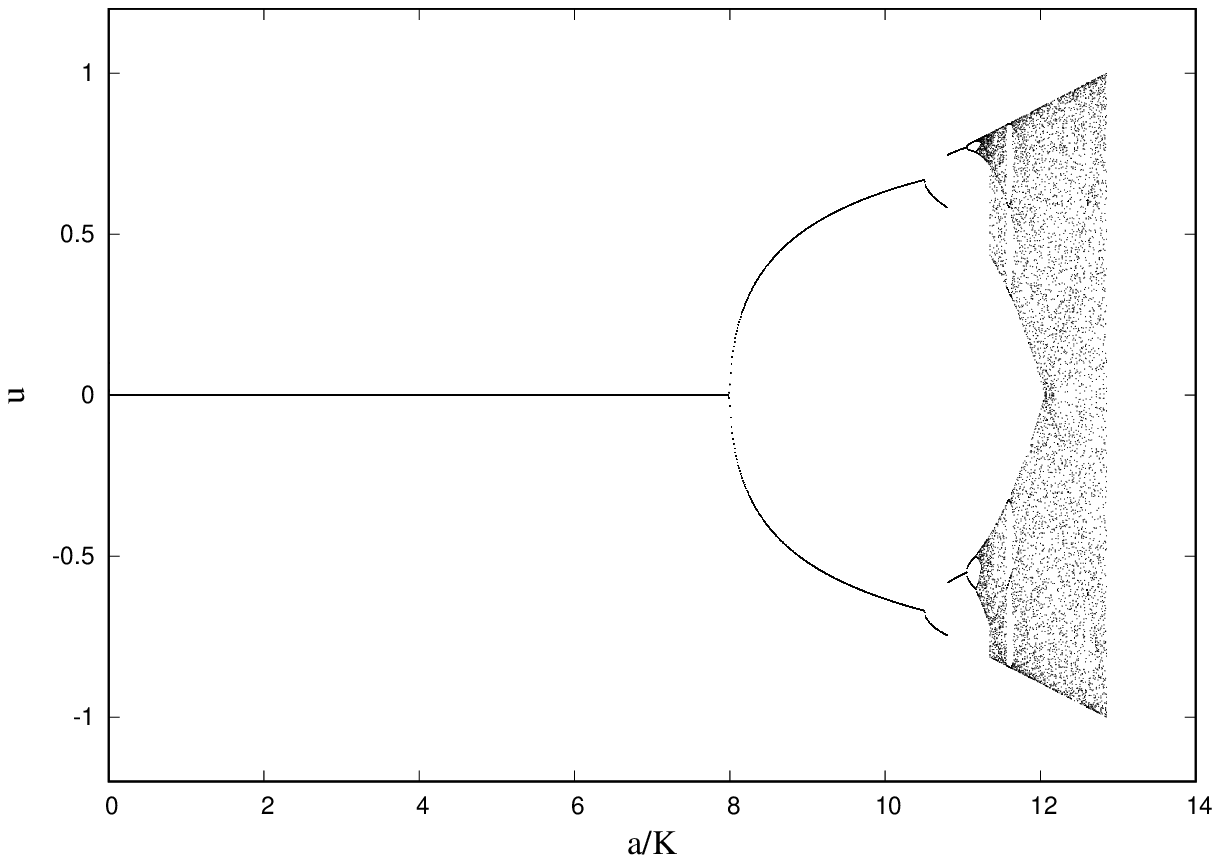}(d) 
\end{minipage}
\caption{\label{fbifa}(Color online) Bifurcation diagram in $\tilde{a}$, for the DKDW map with variable barrier and fixed minima: $\mu=0$ (a), $\mu=0.2$ (b), $\mu=0.5$ (d), $\mu=1$ (d).} 
 \end{figure}
For this first map we also simulated the bifurcation diagram with respect to the deformability parameter $\mu$, considering four different values of $\tilde{a}$ chosen around the first pitchfork bifurcation. Fig. \ref{fbifb} shows that the bifurcation diagram in $\mu$ is dominated by period-halving bifurcations when $\tilde{a}=3.5$, 4, 5 and 6. These period-halving bifurcations are surely indicative of the fact that an increase in the deformability parameter $\mu$ is expected to delay priod-doubling transitions with respect to $\tilde{a}$, as we observed in fig. \ref{fbifa}. 
\begin{figure}\centering
\begin{minipage}{0.5\textwidth}
\includegraphics[width=2.8in,height=2.2in]{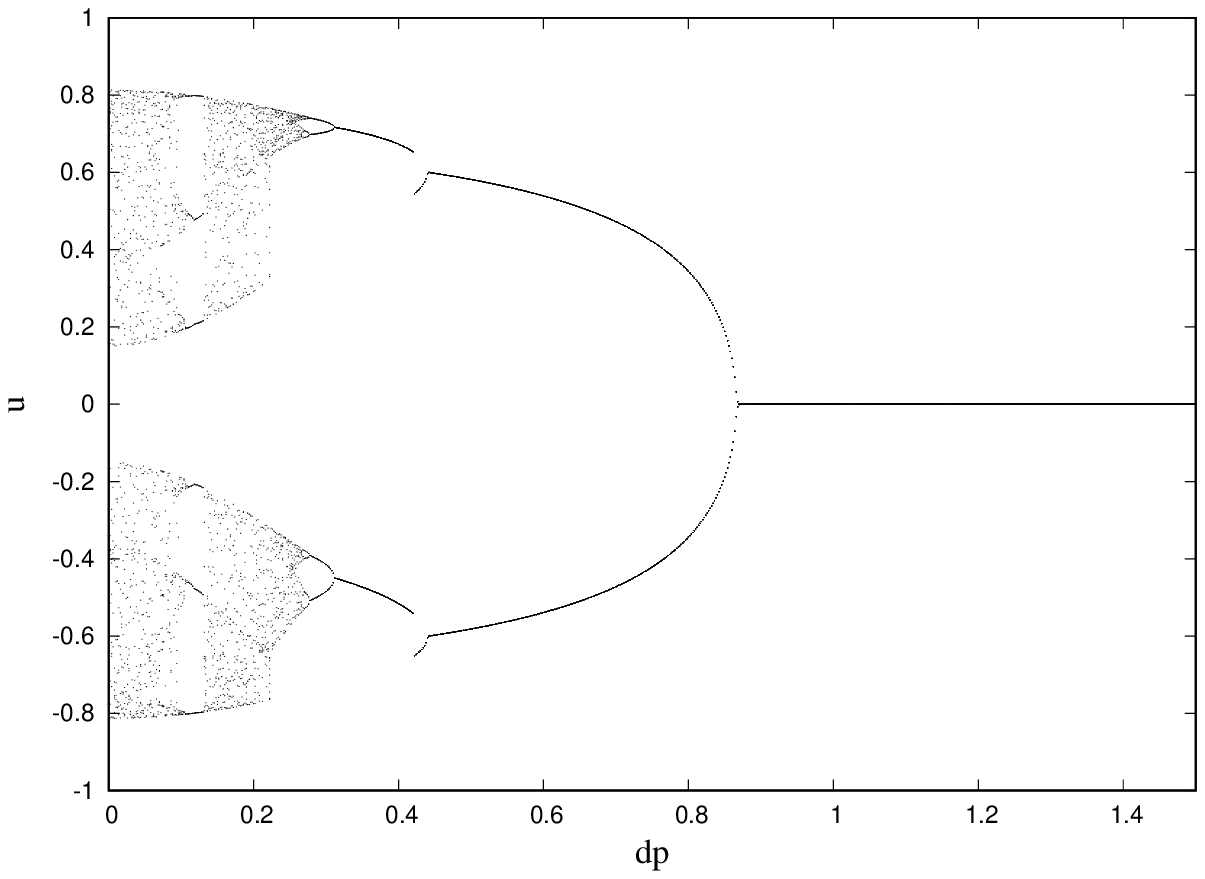}(a) 
\end{minipage}%
\begin{minipage}{0.5\textwidth}
\includegraphics[width=2.8in,height=2.2in]{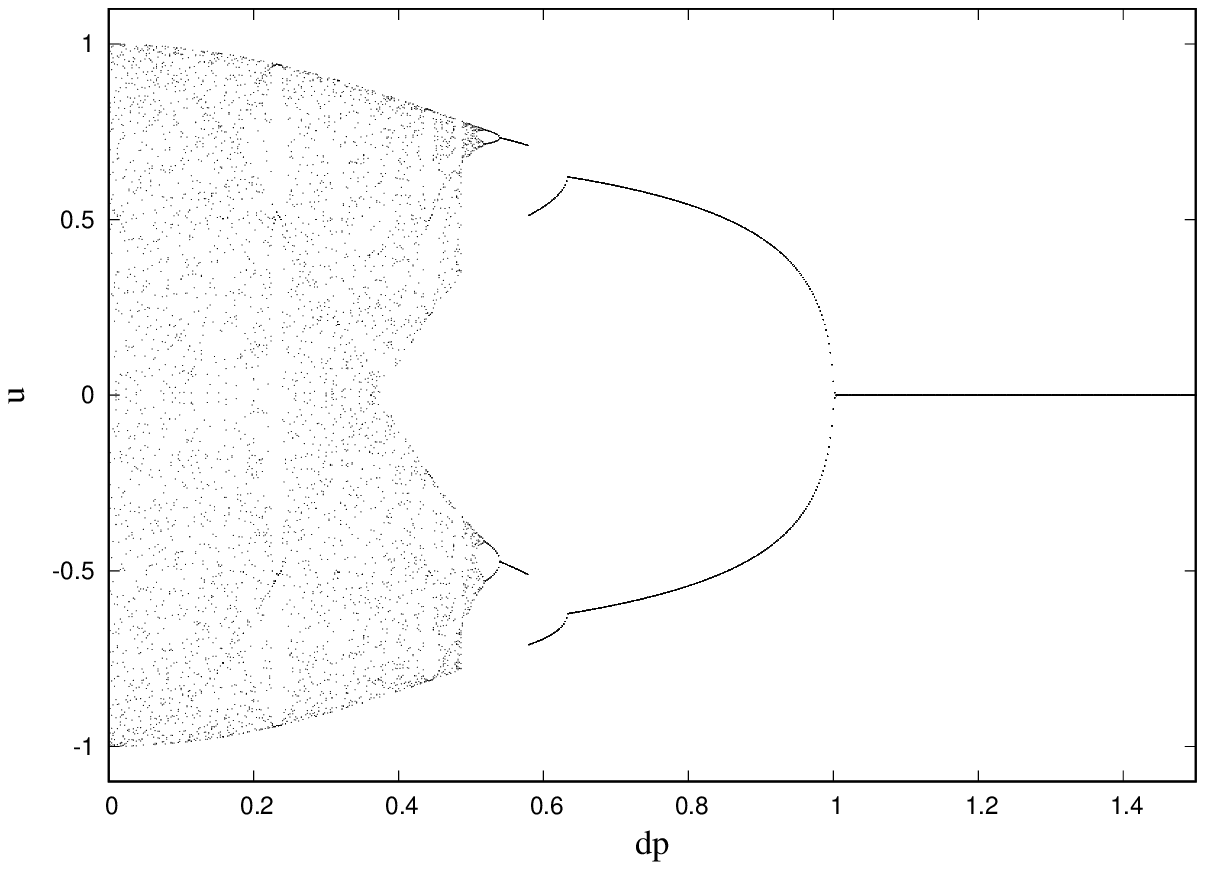}(b) 
\end{minipage}\vskip 0.8truecm
\begin{minipage}{0.5\textwidth}
\includegraphics[width=2.8in,height=2.2in]{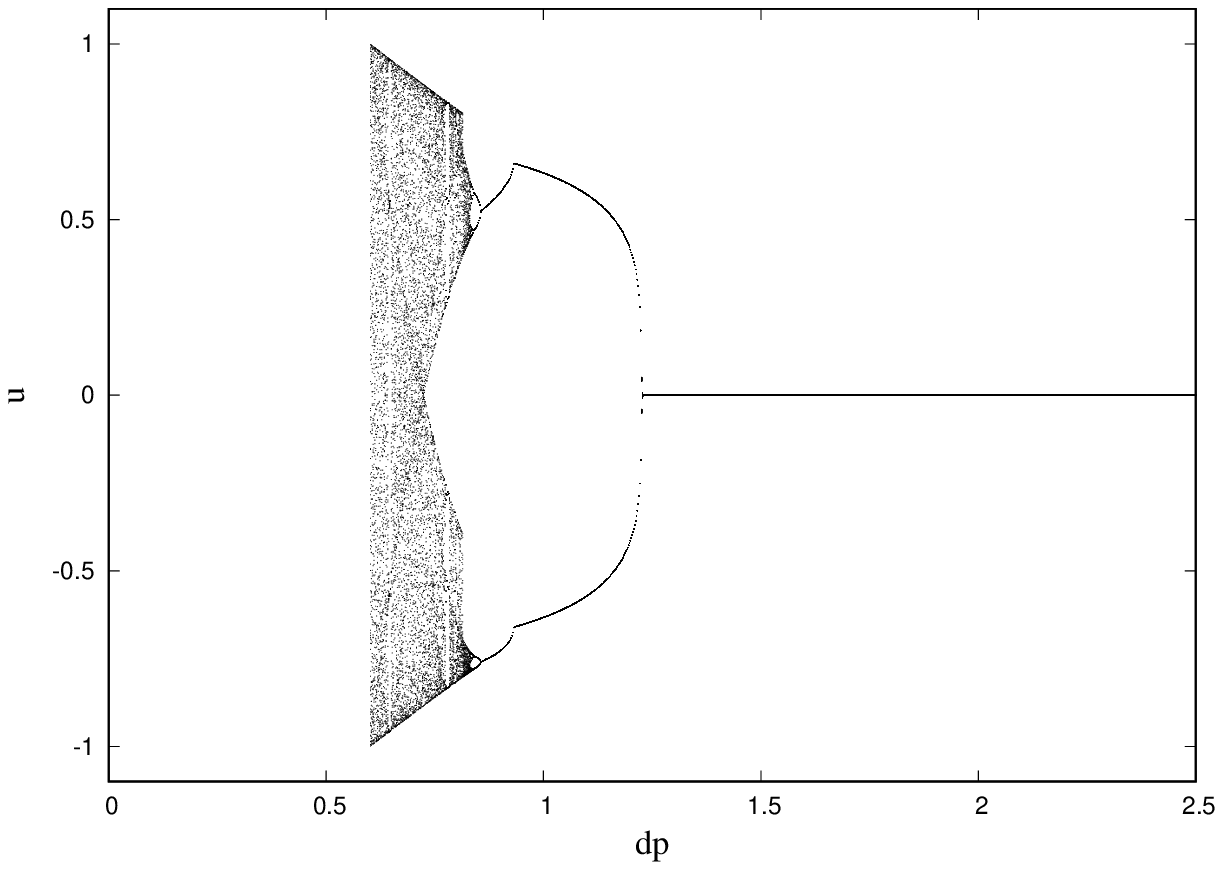}(c) 
\end{minipage}%
\begin{minipage}{0.5\textwidth}
\includegraphics[width=2.8in,height=2.2in]{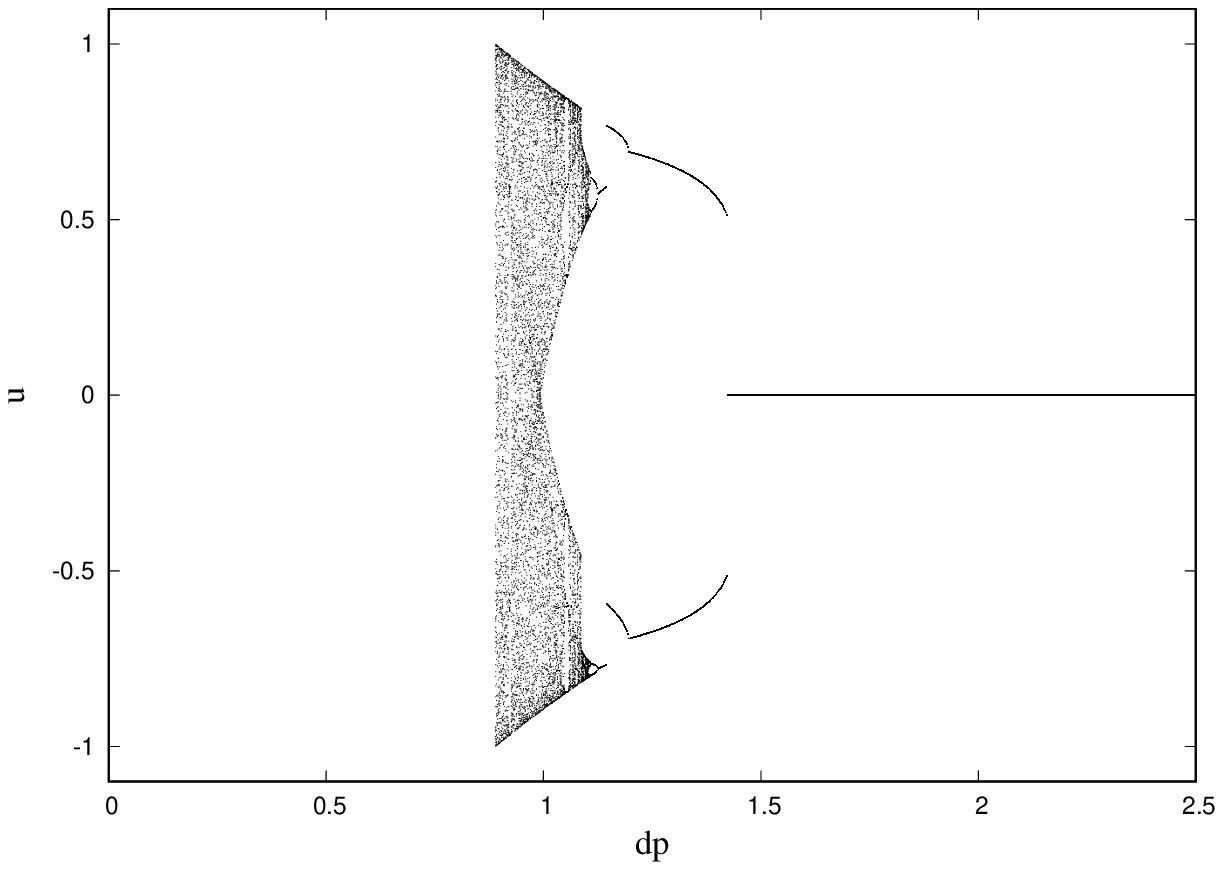}(d) 
\end{minipage}
\caption{\label{fbifb}(Color online) Bifurcation diagram in the deformability parameter (dp) $\mu$, for the DKDW map with variable barrier and fixed minima: (a) $\tilde{a}=3.5$, (b) $\tilde{a}=4$, (c) $\tilde{a}=5$, (d) $\tilde{a}=6$.} 
 \end{figure}
\par The bifurcation diagrams with respect to $\tilde{a}$ for the DKDW map with variable minima are shown in figs. \ref{fbifc1} for the same four different values of $\mu$ used in fig. (\ref{fbifa}. Structures of the bifurcation diagram with respect to $\tilde{a}$ for larger values of $\mu$ (i.e. $\mu=1.5$ for left graph, and 1.8 for right graph), are shown in fig. \ref{fbifc1}.
\begin{figure}\centering
\begin{minipage}{0.5\textwidth}
\includegraphics[width=2.8in,height=2.2in]{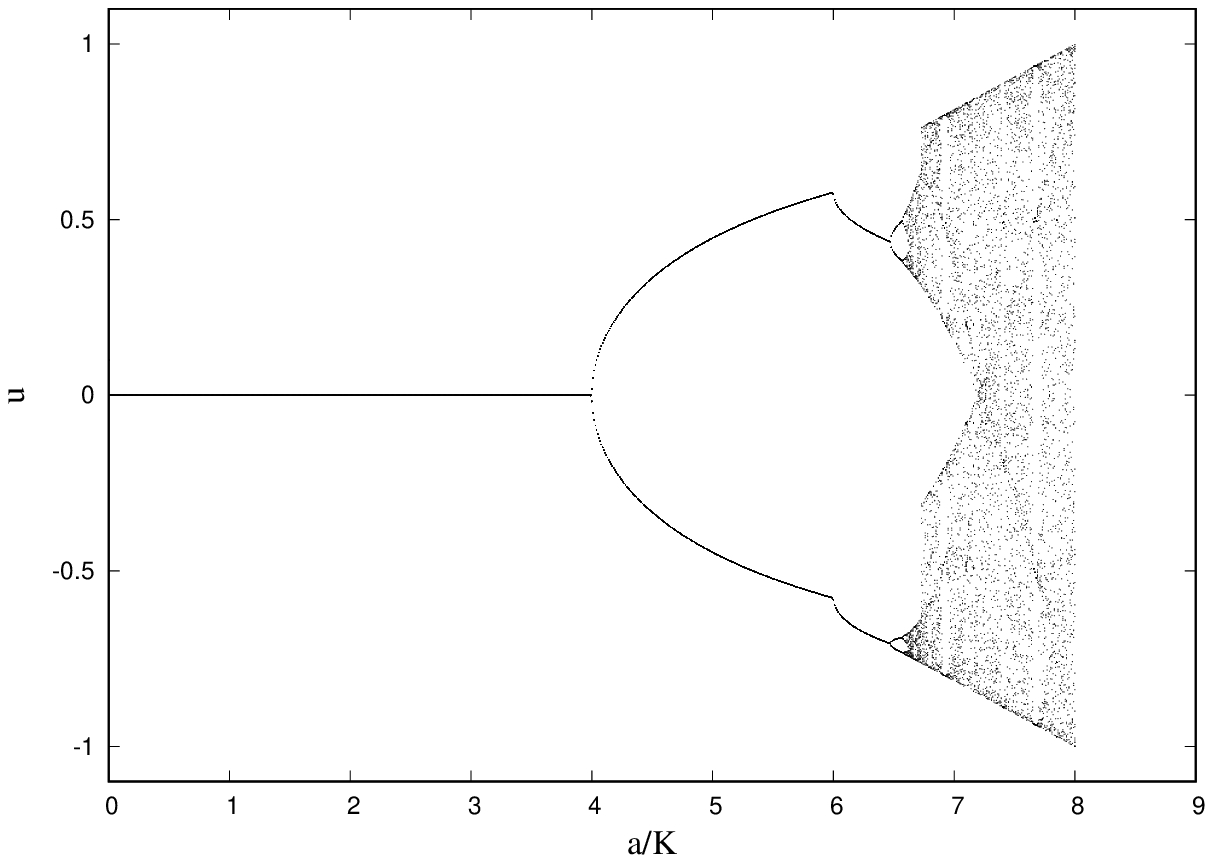}(a) 
\end{minipage}%
\begin{minipage}{0.5\textwidth}
\includegraphics[width=2.8in,height=2.2in]{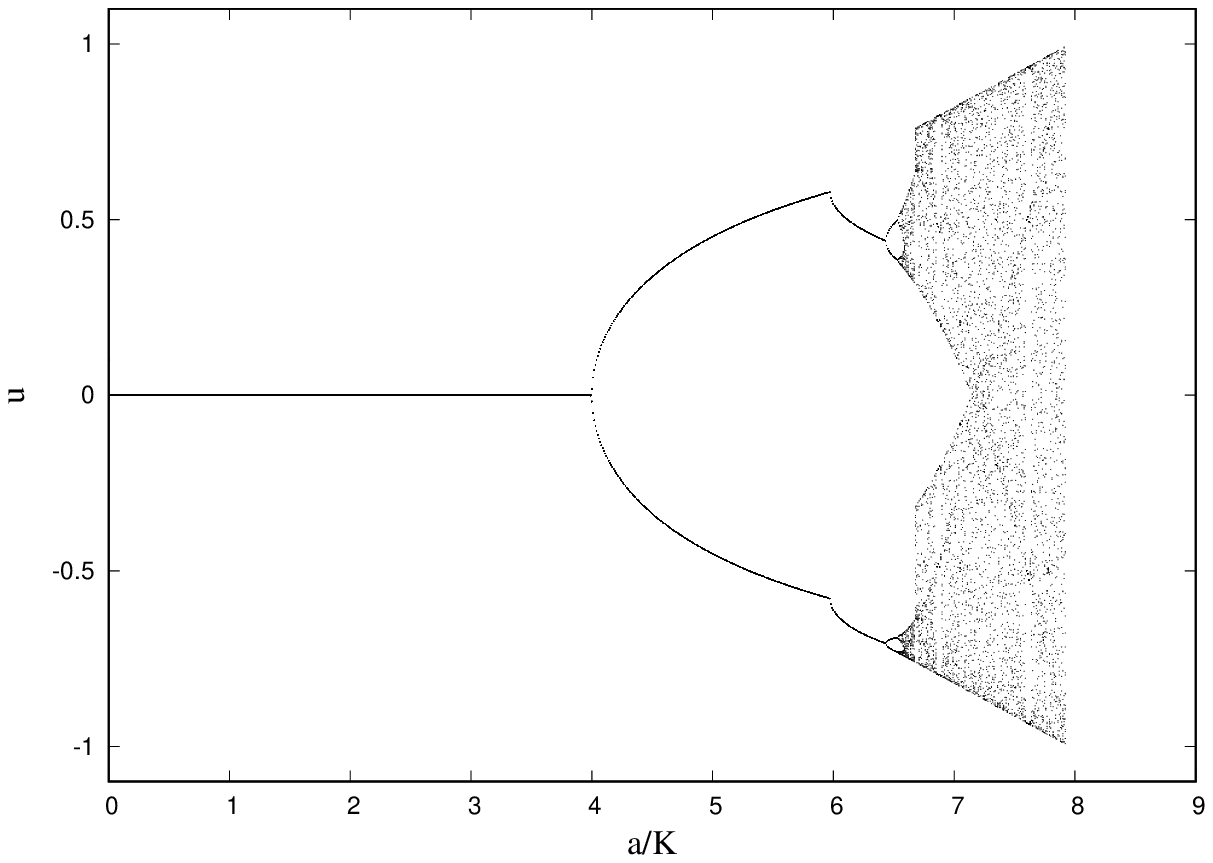}(b) 
\end{minipage}\vskip 0.8truecm
\begin{minipage}{0.5\textwidth}
\includegraphics[width=2.8in,height=2.2in]{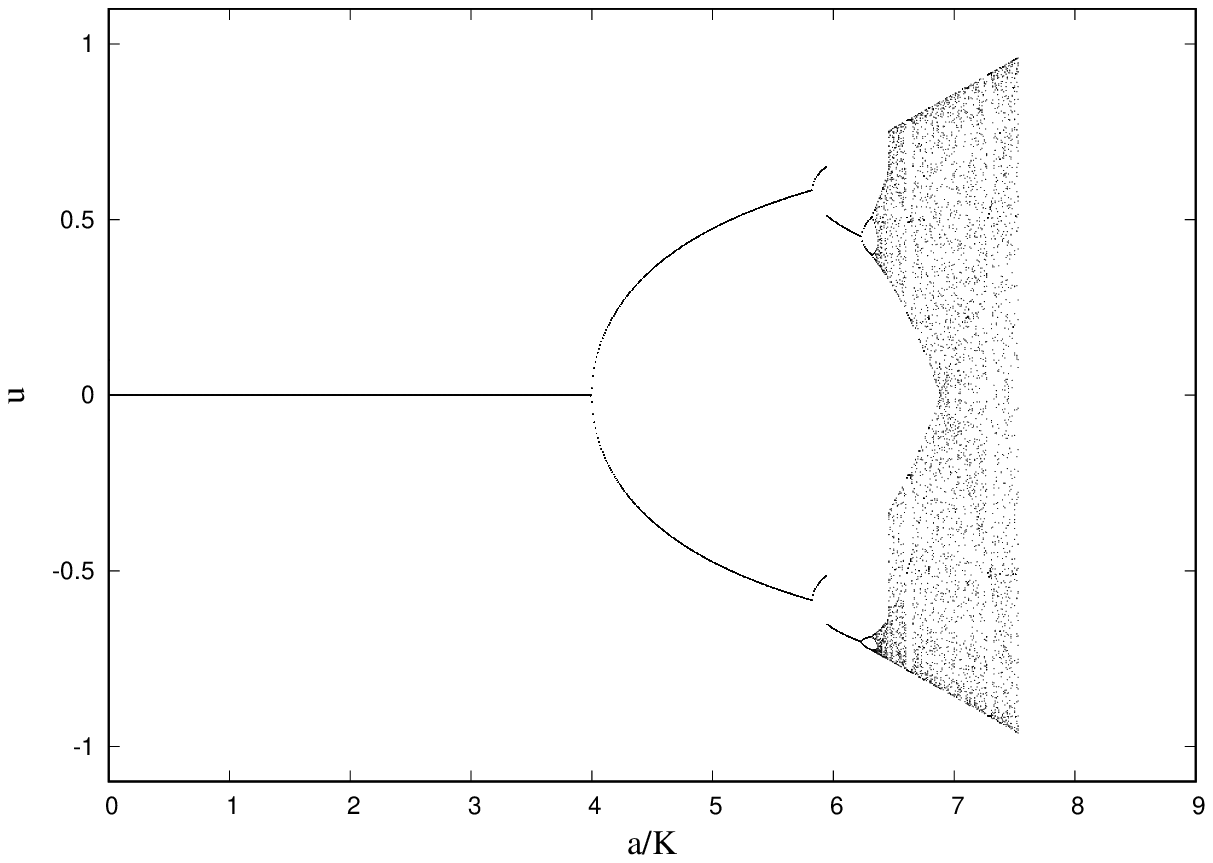}(c) 
\end{minipage}%
\begin{minipage}{0.5\textwidth}
\includegraphics[width=2.8in,height=2.2in]{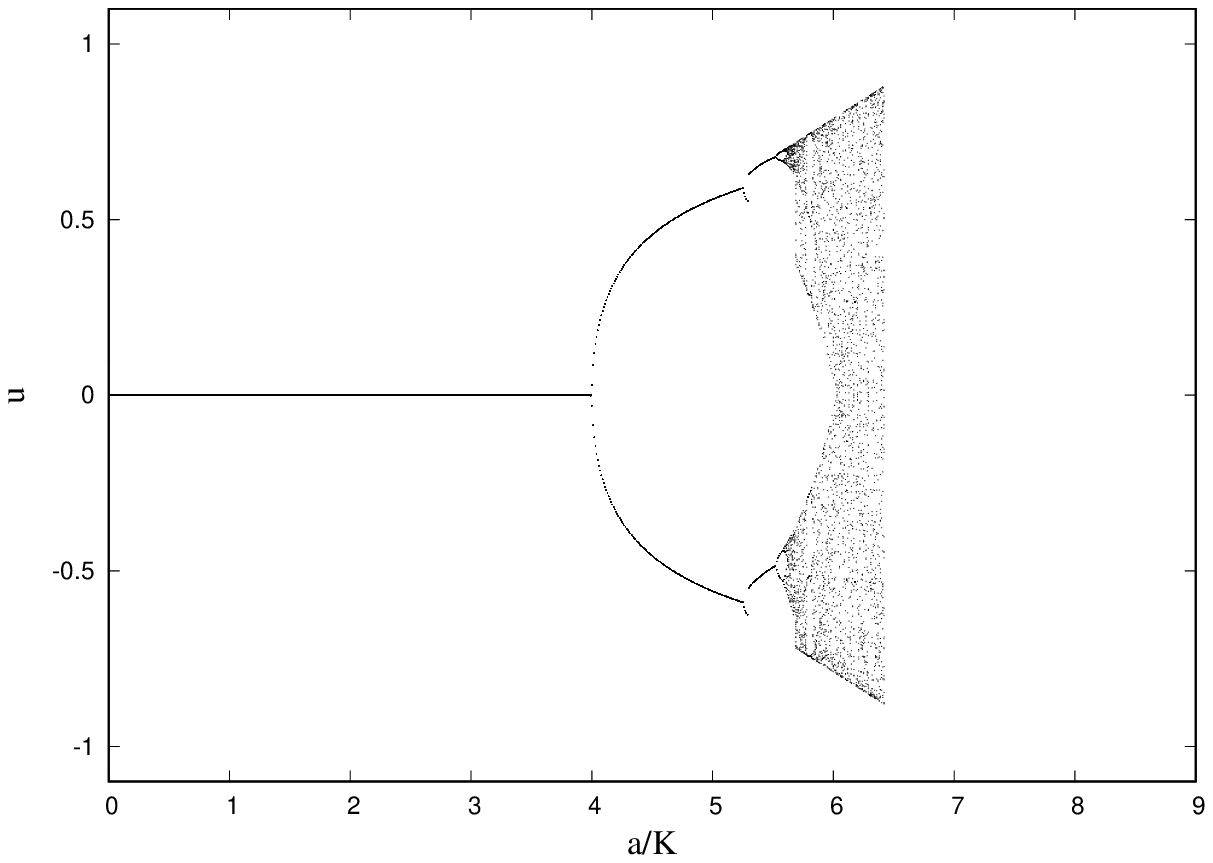}(d) 
\end{minipage}
\caption{\label{fbifc1}(Color online) Bifurcation diagram in $\tilde{a}$, for the DKDW map with variable minima and fixed barrier: (a) $\mu=0$, (b) $\mu=0.2$, (c) $\mu=0.5$, (d) $\mu=1$.} 
 \end{figure}

\begin{figure*}\centering
\begin{minipage}{0.5\textwidth}
\includegraphics[width=2.8in,height=2.2in]{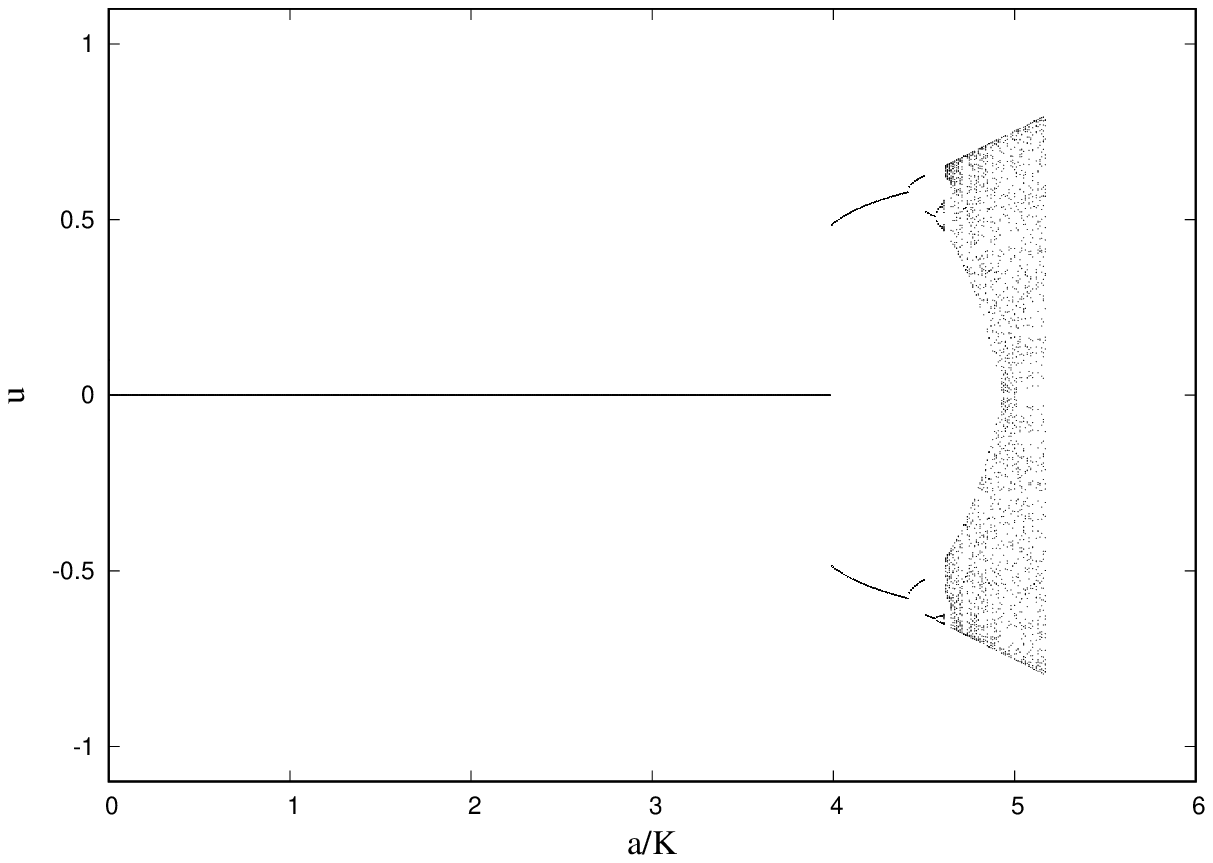}
\end{minipage}%
\begin{minipage}{0.5\textwidth}
\includegraphics[width=2.8in,height=2.2in]{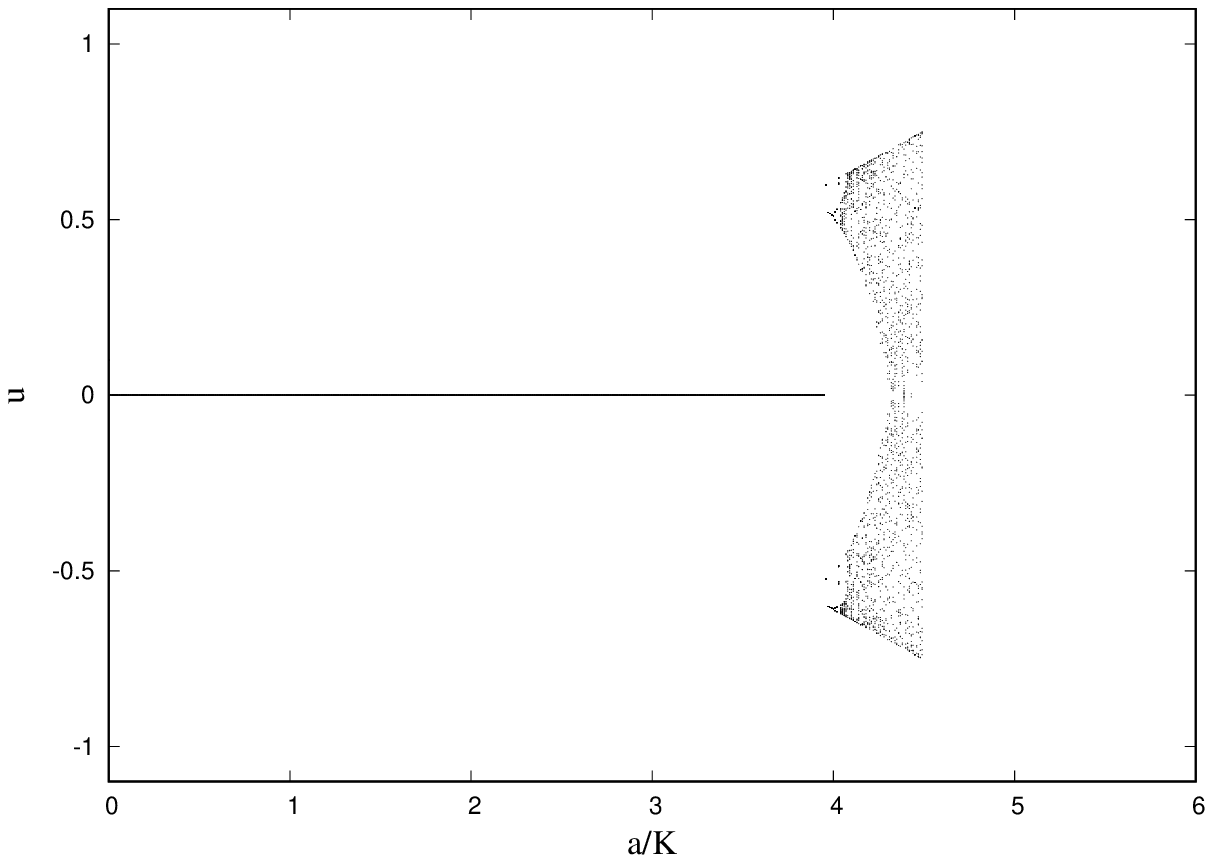}
\end{minipage}
\caption{\label{figam}(Color online)  Bifurcation diagram in $\tilde{a}$, for the DKDW map with variable minima and fixed barrier: $\mu=1.5$ (left panel), and $\mu=1.8$ (right panel).} 
 \end{figure*}
 The bifurcation diagrams with respect to $\mu$ for this second map are plotted in fig. \ref{fbifd}, for four different values of $\tilde{a}$.
 \begin{figure}\centering
\begin{minipage}{0.5\textwidth}
\includegraphics[width=2.8in,height=2.2in]{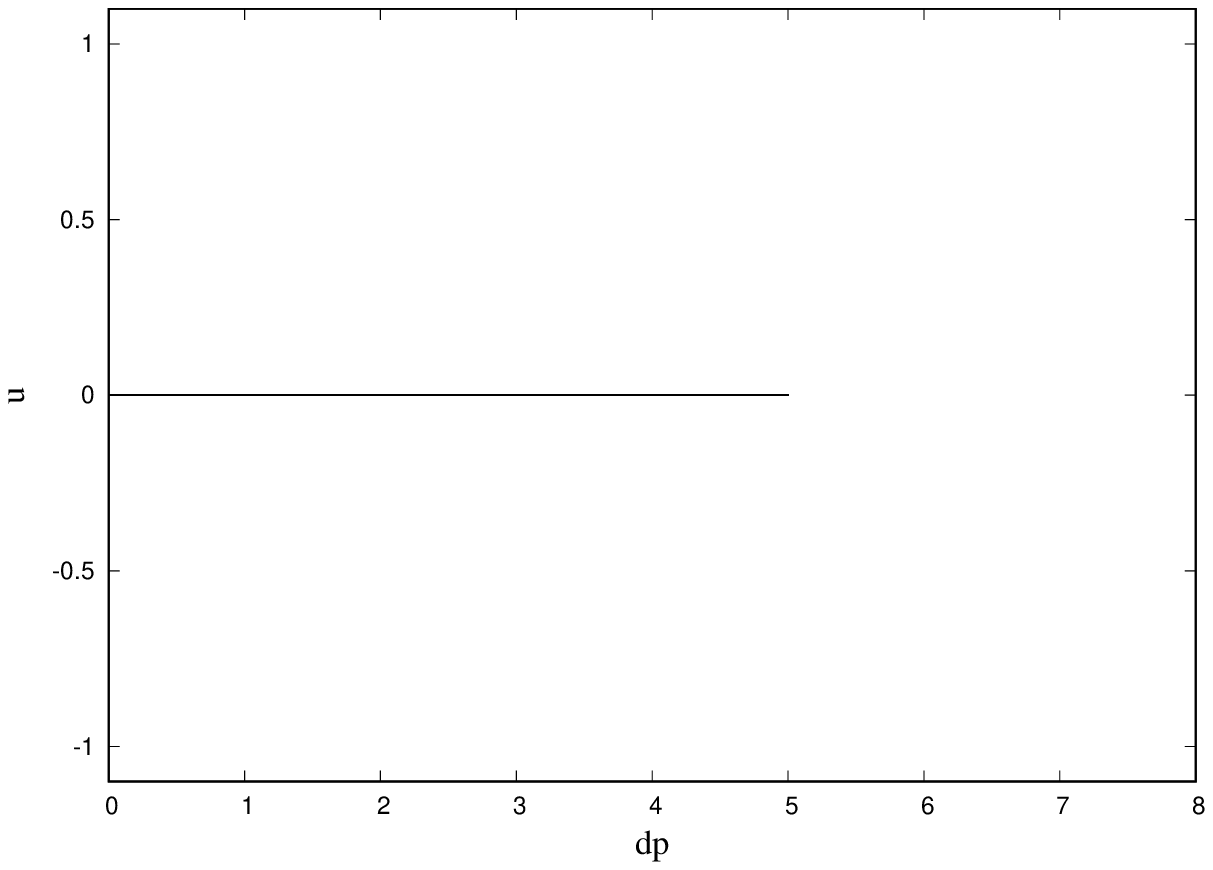}(a) 
\end{minipage}%
\begin{minipage}{0.5\textwidth}
\includegraphics[width=2.8in,height=2.2in]{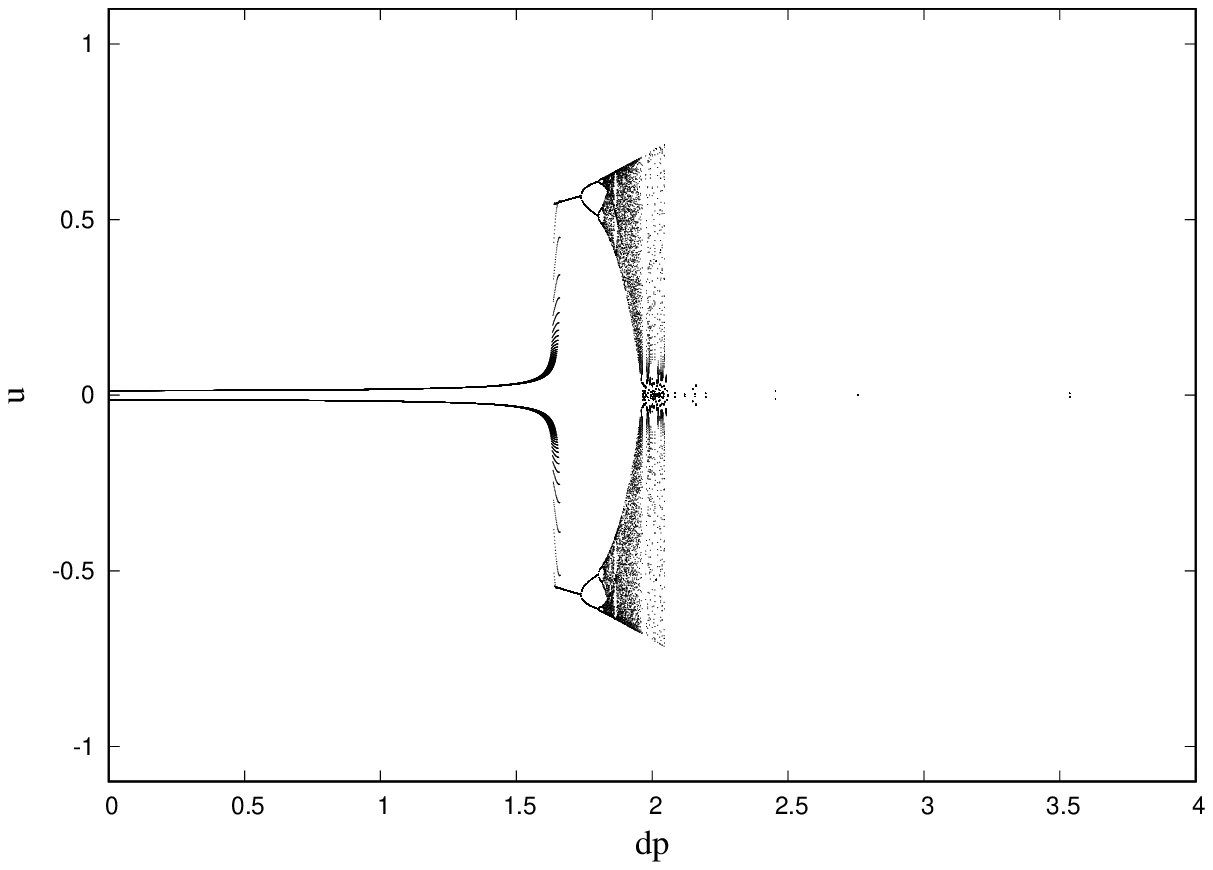}(b) 
\end{minipage}\vskip 0.8truecm
\begin{minipage}{0.5\textwidth}
\includegraphics[width=2.8in,height=2.2in]{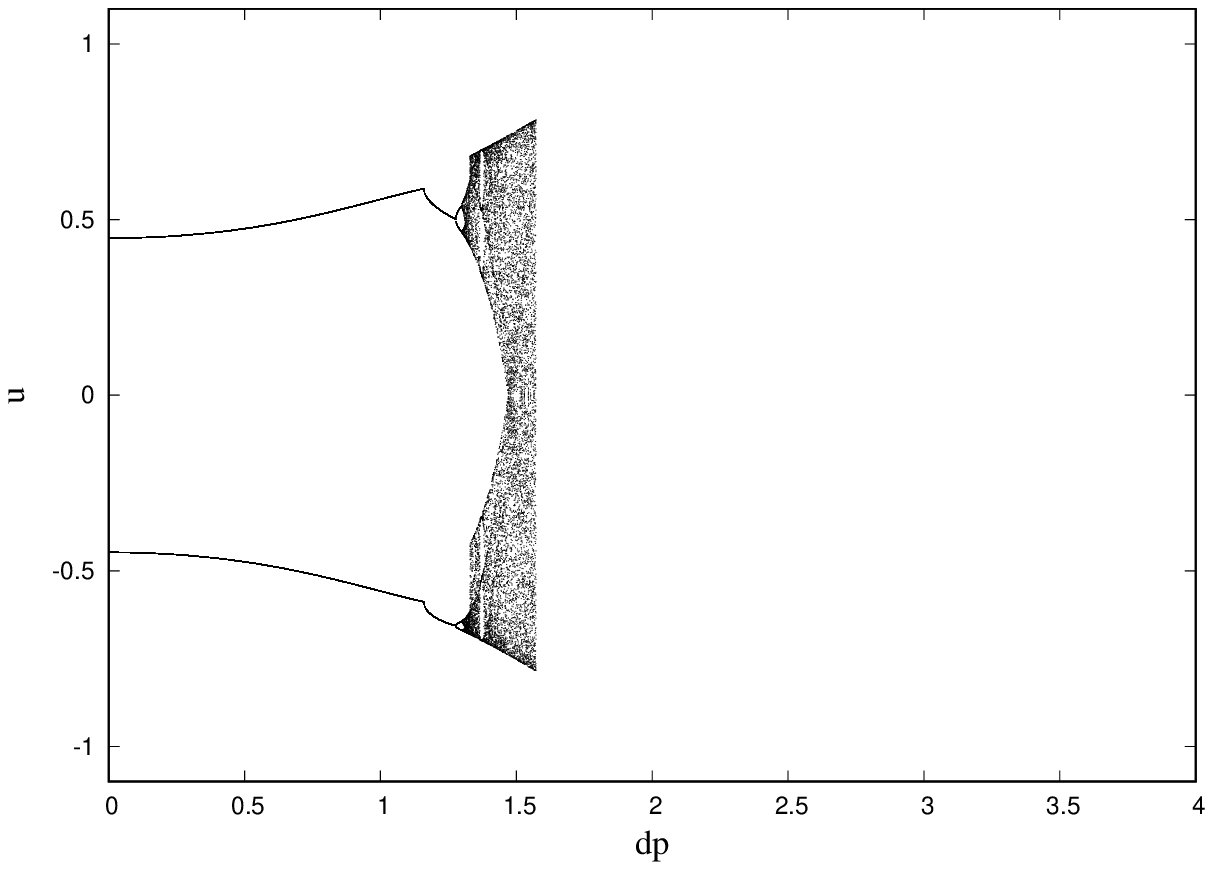}(c) 
\end{minipage}%
\begin{minipage}{0.5\textwidth}
\includegraphics[width=2.8in,height=2.2in]{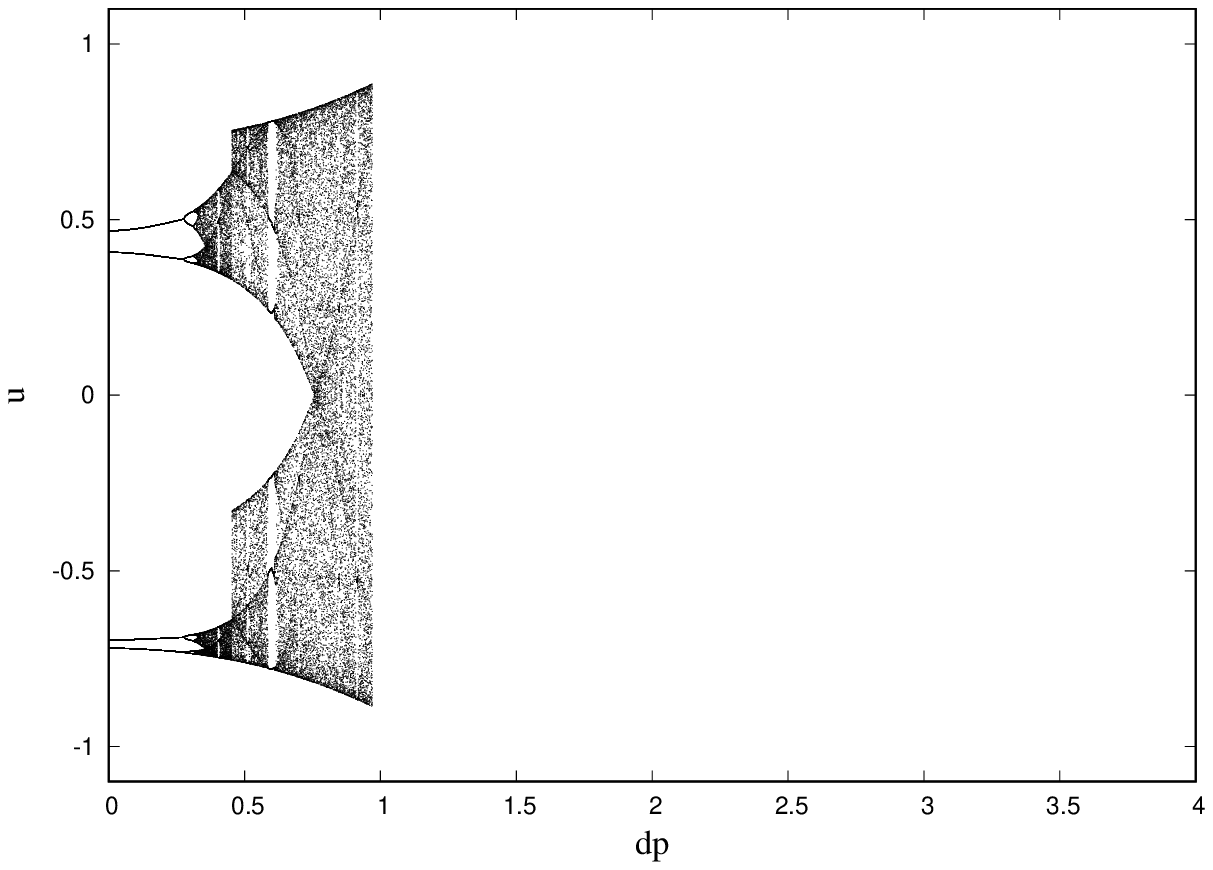}(d) 
\end{minipage}
\caption{\label{fbifd}(Color online) Bifurcation diagram in the deformability parameter (dp) $\mu$, for the DKDW map with variable minima and fixed barrier height: (a) $\tilde{a}=3.9$, (b) $\tilde{a}=4$, (c) $\tilde{a}=5$, (d) $\tilde{a}=6.5$.} 
 \end{figure}
The bifurcation structures emerging from figs. \ref{fbifc1}, \ref{figam} and \ref{fbifd} are all consistent with our previous comments regarding the texture of phase space as well as the analytical prediction on the the first and sexcond bifurcation cascades for the second map. In particular the figures clearly sgow that for the second map, bifurcation diagrams with respect to the two characteristic parameters are dominated by period-doubling cascades. Here the first bifurcation in $\tilde{a}$ is clearly insensitive to the variation of $\mu$, however the second bifurcations leading to period-four orbits occur for values of $\tilde{a}$ that are shifted backward more and more as $\mu$ increases. Hence and increase of $\mu$ shrinks the regions between successive period-doubling bifurcation cascades, so we can conclude that as we increase the values of $\mu$ the difference in values of $\tilde{a}$ at two consecutive bifurcation cascades will decrease. So to say the Feigenbaum number sequence for the DKDW map is non universal, the sequence will strongly depend on values of $\mu$.

\section{\label{sec:level4}Conclusion}
Most physical systems evolve in environments where they are subjected to multiple equilibria of distinct stabilities. Of these, bistable systems form a special class characterized by two energetically equivalent metastable states and eventually an unstable state where the system undergo structural changes. Bistability is a quite common feature in biology, and there are many examples of systems which can operate stably in two completely distinct modes of equivalent symmetries. This is for instance the case for the well known lactose operon in the bacteria Escherichia coli \cite{bis1}, or the phage $\lambda$virus which can exist in either of two states: Under normal conditions this virus can exist in a lysogenic state and survive indefinitely within its host (Escherichia coli).  However, under adverse conditions (such as ultraviolet radiations), the phage can switch to a reproducible (i.e. lytic) mode leading to bacterial lysis \cite{bis2}. Yet another example is the complex system of cross-talking pathways that regulates the decision of cells to enter the process of programmed cell death, also known as apoptosis, as  opposed to continuing normal development \cite{bis2}. These processes in biological systems have established that most often individual cells only exist in one of two distinct states, and upon stimulations the cells undergo a change of state (or a transition) from one state to another. \par
To understand how bistable systems in biology as well as other physical systems such as solid-state materials, biochemical systems and so on, perform complex functions, several mathematical models have been proposed. These models have in common the existence of an  appropriate feedback describing their bistability, and represented by a function with two degenerate minima. This double-well function, which stands for the characteristic energy associated with the appropriate bistable feedback process, is most often represented by the so-called $\phi^4$ potential. For instance the feedback force found in the Duffing model \cite{chua} with soft anharmonicity originates from a $\phi^4$-like double-well potential, this is equally the case in the logistic model with two mestable equilibria separated by a wall, etc. However the rigid profil of the double-well potential representing the bistable feedback function in these mathematical models, is a weakness for it does not enable one to take into consideration the many possible changes in characteristic features of systems as a result of the variability of constraints of their complex living environments. As improvement to this rigidity parametrized double-well potentials were proposed, they include the double-Morse potential \cite{kon1}, the Schmidt potential \cite{schm}, the Magyari motential \cite{mag1}, the razavy potential \cite{raz} and more recently, the family of parametrized potentials whose double-well shapes can be tuned at wish by varying a deformability parameter \cite{dik2}. \\
In the present study we explored the route to chaos for bistable systems described by a discrete $2D$ parametric map, with emphasis on two members of the family of parametreized double-well potentials proposed in ref. \cite{dik3}. We examined the texture of phase space of the two parametric maps as well as the structure of the associate bifurcation diagrams. One of the two discrete $2D$ parametric maps is characterized by a double-well potential whose barrier height can be tuned continuously, leaving unaffected its two degenerate minima. For this first map we obtained that the shape deformability allows shifting to higher magnitudes the values of the ratio $a/K$ where bifurcations are expected to occur. In particular we found that the bifurcation diagram in the deformability parameter was dominated by period-halving cascades, suggesting unambiguously the possible control of route to chaos as well as a non-universal character of the Feigenbaum number sequence for this map. The second map studied has a fixed barrier height but tunable positions of its two degenerate minima. For this second mpa, we found that the first bifurcation always coincides with the first pitchfork instability of the $\phi^4$ map irrespective of the deformability parameter. However, an incease of the deformability parameter seemed to contract the region between consecutive period-douboing cascades. This behavior too, clearly indicates a non-n=universak character of the Feigenbaum number sequences for this second map. \\ 
As we indicated in the introduction, the deformability can be associated with the variability of the living environment of cells and chemical species, as for instance the change of pressures or characteristic parameters of intermocular interactions due to isotopic substitutions (changes of reactants in processes involving chemical reactions), or the change of cell characteristics in the bistability feedback, in reponse to temperature variations \cite{bis3}. The present study particularly points out the fact that while the shape deformability for the models considered here does not change qualitatively intrinsic properties of bistable systems, quantitative changes can however be observed and marked by non-universalities of characteristic features of their route to chaos as for instance the number sequence associated with the recurrence of period-doubling cascades leading to chaos.


\section*{References}
\begin{thebibliography}{}
\bibitem{toda} Toda M 1989 {\it Theory of nonlinear lattices} (Springer Series in Solid-State Sciences, Springer, Berlin).
\bibitem{remoi}Remoissenet M 1999 {\it Waves Called Solitons: Concepts and Experiments} (Springer, Berlin).
\bibitem{chua}Chua L 1999 {\it Visions of Nonlinear Science in the $21st$ Century} (Nonlinear Science Series vol. 26, World Scientific, Singapore).
\bibitem{pana}Chong C and Kevrekidis P G 2018 {\it coherent structures in granular crystals: from experiment and modelling to computation and mathematical analysis} (Springer, Berlin).
\bibitem{sin2}Currie J F, Krumhansl J A, Bishop A R and Trullinger S E 1980 {\it Phys. Rev.} B{\bf 22}, 477.
\bibitem{sin3}Tabor M 1989 {\it The Sine-Gordon Equation} in {\it Chaos and Integrability in Nonlinear Dynamics: An Introduction} (Wiley, New York) p 305. 
\bibitem{fren}Braun O M and Kivshar Y S 2004 {\it The Frenkel-Kontorova Model: Concepts, Methods and Applications} (Springer, New York).   
\bibitem{dikan}Dikand\'e A M and Kofan\'e T C 1995 {\it J. Phys.: Condens. Mat.} {\bf 7}, L141.
\bibitem{lcrys}Dikand\'e A M, Nyanga B Y
and Mkam Tchouobiap S E 2018 {\it Mod. Phys. Lett.} B {\bf 32}, 1850392. 
\bibitem{com}Dikand\'e A M and Kofan\'e T C 1994 {\it Phys. Scrip.} {\bf 49}, 110.
\bibitem{kiv}Kivshar Y S and Malomed B M 1989 {\it Rev. Mod. Phys.} {\bf 61}, 763.
\bibitem{krum}Krumhansl J A and Schrieffer J R 1975 {\it Phys. Rev.} B {\bf 11}, 3535.  
 \bibitem{bak2}Bak P and Pokrovsky V L 1981 {\it Phys. Rev. Lett} {\bf 47}, 958.  
\bibitem{bak1}Bak P and Jensen M H 1982 {\it J. Phys. A} {\bf 15}, 1893. 
\bibitem{fer1}Blinc R and Zeks B 1974 {\it Soft modes in ferroelectrics and antiferroelectrics} (North-Holland, Amsterdam.
\bibitem{fer2}Riste T 1981 {\it Nonlinear Phenomena at Phase Transitions and Instabilities} (Plenum, New York).
\bibitem{bio1}Martinez-Corral R, Liu J, S\"uel G M and Garcia-Ojalvo J 2018 {\it PNAS} {\bf 36}, 115.
\bibitem{bio2}Mosekilde E 1998 {\it Topics in nonlinear dynamics: Applications to Physics, Biology and Economic Systems} (World Scientific Publishing, Singapore).  
\bibitem{econ1}Alatriste-Contreras M G, Brida J G and Anyul M P 2019 {\it J. Dyn. Games} {\bf 6}, 87.
\bibitem{cros1}Kalmykov Y P, Coffey W T and Titov S V 2006 {\it J. Chem. Phys.} {\bf 124}, 024107.
\bibitem{cros2}Kumar Ghosh P, Barik D, Bag B C and Shankar Ray D 2006 {\it J. Chem. Phys.} {\bf 123}, 224104.
\bibitem{crosb}Ushiyama H and Takatsuka H 2001 {\it J. Chem. Phys.} {\bf 115}, 5903.
\bibitem{cros3}McCoy A B, Huang X, Carter S and Bowman J M 2005 {\it J. Chem. Phys.} {\bf 123}, 064317.
\bibitem{cros3a}Bierman A 1966  {\it J. Chem. Phys.} {\bf 45}, 647.
\bibitem{cros4}Barrow G M 1957 {\it J. Chem. Phys.} {\bf 26}, 558.
\bibitem{mag1}Magyari E 1981 {\it Z. Phys.} B{\bf 43}, 345.
\bibitem{schm}Schmidt V H 1979 {\it Phys. Rev.} B {\bf 20}, 4397.
\bibitem{kon}Konwent H, Machnikowski O and Radosz A 1996 {\it J. Phys.: Condens. Matt.} {\bf 8}, 4325.
\bibitem{kon1}Konwent H, Machnikowski P, Magnuszewski P and Radosz A 1998 {\it J. Phys. A: Math. Gen.} {\bf 31}, 7541.
\bibitem{raz}Razavy M 1980 {\it Am. J. Phys.} {\bf 48}, 285.
\bibitem{dik1}Dikand\'e A M and Kofan\'e T C 1991 {\it J. Phys.: Condens. Matt.} {\bf 3}, L5203.    
\bibitem{dik2}Kofan\'e T C and Dikand\'e A M 1993 {\it Solid State Commun.} {\bf 86}, 749.   
\bibitem{dik3}Dikand\'e A M and Kofan\'e T C 1994 {\it Solid State Commun.} {\bf 89}, 283. 
\bibitem{dik4}Dikand\'e A M and Kofan\'e T C 1994 {\it Solid State Commun.} {\bf 89}, 559.  
\bibitem{zh1}Zhou J B, Liang J Q and Pu F C 2001 {\it Phys Lett} A. {\bf 278}, 243.
\bibitem{zh2}Bazeia D, Gomes A R, Nobrega K Z and Simas F C 2020 {\it Phys. Lett.} B {\bf 803}, 135291.
\bibitem{zh3}Dikand\'e A M 2002 {\it Phys. Lett.} A {\bf 304}, 143.
\bibitem{zh4}Naha Nzoupe F, Dikand\'e A M and Tchawoua C 2020 {\it Math. Methods Appl. Sci.}, 1-15. DOI: https://doi.org/10.1002/mma.6965.
\bibitem{bak3}Jensen M H, Bak P and Popielewicz A 1983 {\it J. Phys. A: Math. Gen.} {\bf 16}, 4369.
\bibitem{zh5}Dikand\'e A M and Epie Njumbe E 2010 {\it Phys. Scrip.} {\bf 81}, 055002.
\bibitem{zh6}Dikand\'e A M, Epie Njumbe E and Kofan\'e T C 2008 {\it Phys. Lett.} A {\bf 372}, 6890.
\bibitem{quisp}Quispel G R W, Roberts J A G and Thompson J C 1989 {\it Physica} D {\bf 34}, 183.
\bibitem{quisp2}McMillan E M 1979 {\it Topics in Modern Physics} (Ed. by W. E. Brittin and H. Odabasi, Boulder: Colorado University) p 219.
\bibitem{quisp3}Quispel G R W, Roberts J A G and Thompson C J 1988 {\it Phys. Lett.} A {\bf 126}, 419.
\bibitem{pa}Samara, G A 1978 {\it Ferroelec.} {\bf 20}, 87.
\bibitem{p1}Katrusiak A and Szafra\'nski A 1999 {\it Phys. Rev. Lett.} {\bf 82}, 576.
\bibitem{p2}Tomaszewski P 1992 {\it Phase Trans.} {\bf 38}, 127.
\bibitem{p3}Ye H Y, Fu D W, Zhang Y, Zhang W, Xiong R G and Huang S D 2009 {\it J. Am. Chem. Soc.} {\bf 131}, 42.
\bibitem{p4}Szafra\'nski M, Katrusiak S and McIntyre G J 2002 {\it Phys. Rev. Lett.} {\bf 89}, 215507.
\bibitem{greene1}Greene J M 1979 {\it J. Math. Phys}. {\bf 20}, 1183. 
\bibitem{brent}Brent R P 1973 {\it Algorithms for Minimization Without Derivatives} (Englewood Cliffs, Prentice-Hall, New Jersey). 
\bibitem{bis1}Santill\'an M 2008 {\it J. Biophys.} {\bf 94}, 2065.
\bibitem{bis2}Chaves M, Eissing Th and Allg\"ower F 2008 {\it IEEE Trans. Auto. Control} {\bf 53}, 87.
\bibitem{bis3}Nuss A M et al. {\it A Precise Temperature-Responsive Bistable Switch Controlling Yersinia Virulence}, PLoS Pathog. 12: e1006091. https://doi.org/10.1371/journal.ppat.1006091.

\end{thebibliography}
\end{document}